\documentclass[sigconf,nonacm]{acmart}

\usepackage{subcaption} 
\usepackage{multirow}
\usepackage{graphicx}
\usepackage{siunitx}
\usepackage{natbib}
\usepackage{lipsum}
\usepackage{placeins}
\AtBeginDocument{%
  }

\setcopyright{acmlicensed}
\copyrightyear{2018}
\acmYear{2018}
\acmDOI{XXXXXXX.XXXXXXX}

\acmConference[WWW '26]{The International World Wide Web Conference}{April 13--17, 2026}{Dubai, UAE}
\acmISBN{978-1-4503-XXXX-X/18/06}

\begin{document}

\newcommand{\GG}[1]{\textcolor{red}{[GG: #1]}}
\newcommand{\comments}[1]{\textcolor{red}{[#1]}}

\title{Sparse Autoencoders for Sequential Recommendation Models: Interpretation and Flexible Control}

\author{Anton Klenitskiy}
\email{antklen@gmail.com}
\orcid{0009-0005-8961-6921}
\affiliation{
  \institution{Sber AI Lab}
  \city{Moscow}
  \country{Russian Federation}
}

\author{Konstantin Polev}
\email{endless.dipole@gmail.com}
\affiliation{
  \institution{Sber AI Lab}
  \city{Moscow}
  \country{Russian Federation}
}

\author{Daria Denisova}
\email{duny.explorer@gmail.com}
\affiliation{
  \institution{Sber AI Lab}
  \city{Moscow}
  \country{Russian Federation}
}

\author{Alexey Vasilev}
\email{alexxl.vasilev@yandex.ru}
\orcid{0009-0007-1415-2004}
\affiliation{
  \institution{Sber AI Lab, HSE University}
  \city{Moscow}
  \country{Russian Federation}
}

\author{Dmitry Simakov}
\email{dmitryevsimakov@gmail.com}
\orcid{0009-0003-3199-479X}
\affiliation{
  \institution{Sber AI Lab}
  \city{Moscow}
  \country{Russian Federation}
}

\author{Gleb Gusev}
\email{gleb57@gmail.com}
\affiliation{
  \institution{Sber AI Lab}
  \city{Moscow}
  \country{Russian Federation}
}

\renewcommand{\shortauthors}{Klenitskiy et al.}

\begin{abstract}
  Many current state-of-the-art models for sequential recommendations are based on transformer architectures. Interpretation and explanation of such black box models is an important research question, as a better understanding of their internals can help understand, influence, and control their behavior, which is very important in a variety of real-world applications. Recently, sparse autoencoders (SAE) have been shown to be a promising unsupervised approach to extract interpretable features from neural networks. 

In this work, we extend SAE to sequential recommender systems and propose a framework for interpreting and controlling model representations. We show that this approach can be successfully applied to the transformer trained on a sequential recommendation task: directions learned in such an unsupervised regime turn out to be more interpretable and monosemantic than the original hidden state dimensions.
Further, we demonstrate a straightforward way to effectively and flexibly control the model's behavior, giving developers and users of recommendation systems the ability to adjust their recommendations to various custom scenarios and contexts.

\end{abstract}



\maketitle

\section{Introduction}

Sequential recommender systems capture the dynamic nature of user preferences by taking into account the temporal order of user actions. This approach is particularly valuable in online environments, such as e-commerce platforms, social networks, and content streaming services, where user interests evolve and change rapidly, and the sequence of interactions strongly influences future behavior. By incorporating this sequential information, models can provide more accurate and relevant recommendations, improving personalization and user experience.

Transformer-based models, such as SASRec~\cite{kang2018self}, BERT4Rec~\cite{sun2019bert4rec}, and their numerous improvements~\cite{mezentsev2024scalable,petrov2023gsasrec,tikhonovich2025esasrec,li2020time,frolov2024self}, effectively model complex sequential patterns. Interpretation and explanation of these models are of high practical importance due to their increasing usage both in research and industrial applications. Explainable recommendations, providing transparency into why certain recommendations are made, can help detect biases, improve user trust and the reliability of recommendations. In addition, interpretability may help to adjust recommendations to specific contexts, individual user preferences, and needs, leading to a more satisfying and personalized experience.

Sparse Autoencoders (SAE) have recently emerged as a powerful unsupervised approach for discovering interpretable directions in the latent representations of neural networks. While this technique has been primarily explored in the domains of language and vision~\cite{cunninghamSparseAutoencodersFind2023a,makelovPrincipledEvaluationsSparse2024,gaoScalingEvaluatingSparse2024,gortonMissingCurveDetectors2024}, its application to recommendation models is still at an early stage. Recent work~\cite{wang2024interpretinternalstatesrecommendation} has begun to explore the use of sparse autoencoders for interpretation of recommender systems. We further develop SAE-based interpretation methods and conduct a systematic study of the ability to influence model outputs. Recommendation data is structured around user–item interactions rather than tokens, and items are often associated with rich metadata and attributes. These properties require adapting the SAE pipeline to account for domain-specific structure and to define meaningful interpretability metrics grounded in item attributes.

We extend SAE to sequential recommendations and propose a principled methodology for evaluating and interpreting the learned representations. We introduce quantitative metrics for assessing the interpretability of SAE features based on explicit item attributes, as well as procedures for identifying meaningful directions in the model’s latent space. Beyond interpretability, we show that these directions enable precise and flexible control over model behavior by amplifying or suppressing specific item properties - an effect known as steering~\cite{bricken2023towards}. We perform a systematic analysis demonstrating that steering works reliably in recommendation models and measure its impact on both the semantics and quality of recommendations. Finally, we validate the approach in two distinct domains (music and movies), showing that it generalizes across datasets with different attribute structures and is applicable to a wide range of personalization scenarios.

In short, the main contributions of this paper are:
\begin{itemize}%
    \item We extend the sparse autoencoders to sequential recommendation models, proposing methods to identify interpretable features and  metrics to assess interpretability using item attributes.
    \item We demonstrate that SAE learns numerous interpretable, non-trivial, and semantically distinct features that align with meaningful item properties.
    \item We show that SAE enables flexible control over model predictions and quantitatively evaluate its effectiveness and influence on recommendations.
\end{itemize}

\section{Related Work}

Various approaches have been proposed to interpret recommendation models. In the study \cite{Tsang2020FeatureII}, automatic detection of important feature interactions in neural networks is introduced. Other interpretation techniques are based on language modeling~\cite{Hada2021,li-etal-2021-personalized,li2023}, attention mechanism~\cite{xiao2017attentionalfactorizationmachineslearning,8896879},
counterfactual methods~\cite{Tan2021,Kaffes2021}, and other explanation strategies. While these methods help reveal aspects of the model's internal structure, they are often closely tied to specific architectures, which limits their generalizability.
Recent work~\cite{wang2024interpretinternalstatesrecommendation} applied SAE for interpreting recommendation models, relying on automated interpretation with a language model based on textual descriptions, an approach inherited from NLP. In contrast, we propose a domain-specific methodology that leverages explicit item attributes to interpret learned representations, introduce new quantitative metrics for evaluating interpretability, and provide a comprehensive analysis of model control.

Several methods have also been developed to enable controllable recommendations. One approach~\cite{cen2020controllablemultiinterestframeworkrecommendation} identifies user interests through clustering and generates recommendations for each cluster with adjusted weights, allowing for more targeted suggestions. Another line of work focuses on using text to guide recommendations. In \cite{Mysore2023}, a textual user profile is constructed from interaction history, and modifying this profile allows for generating recommendations with desired characteristics. LLMs have also been applied for interactive recommendation control~\cite{lu-etal-2024-aligning} by customizing them using instructions or contextual descriptions. However, language models may not always incorporate user preferences in a predictable or reliable way.

Some recent works have also applied sparse autoencoders to recommender systems, but with different objectives. In~\cite{kasalicky2025future}, SAE is used primarily as a compression mechanism, projecting dense embeddings into a high-dimensional, sparsely activated space. Another study~\cite{ahmadov2025opening} employs SAE to identify neurons encoding popularity-related signals and adjust their activations to mitigate popularity bias in a sequential recommendation model.

\section{Sparse Autoencoders}
\label{sec:sparse_autoencoders}
SAE has emerged as a tool in the field of LLM mechanistic interpretability through the employment of techniques from sparse dictionary learning that were previously applied to tasks such as understanding neuronal signals in the visual cortex ~\cite{olshausenSparseCodingOvercomplete1997, olshausenSparseCodingSensory2004, cunninghamSparseAutoencodersFind2023a}. Building on experiments with smaller transformer models~\cite{elhageToyModelsSuperposition2022, cunninghamSparseAutoencodersFind2023a}, SAE-based approaches have been further developed and scaled to larger models, including cutting-edge proprietary LLMs. These advancements continue to provide insights into model internals and offer promising strategies for better controlling model generation~\cite{gaoScalingEvaluatingSparse2024}.

Sparse autoencoders (SAEs) are a class of autoencoders designed to produce compact and interpretable latent representations by ensuring that only a small subset of hidden units becomes active for any given input. We consider the classic formulation, where an input \( x \in \mathbb{R}^n \) is mapped to a hidden representation \( h(x) \in \mathbb{R}^m \) through an encoder and then reconstructed by a decoder to approximate the original input, \( \hat{x} \). 
This can be expressed as \( h(x) = ReLU(Wx + b) \) and \( \hat{x} = W'h(x) + b' \), where \( W, W' \) are the weight matrices, \( b, b' \) are bias terms. 
The training objective is to minimize the reconstruction loss, \( \|x - \hat{x}\|_2^2 \) while imposing a sparsity constraint on the hidden representation \( h(x) \).

A common approach to enforce sparsity on SAE is to add an \( L_1 \) regularization term on \( h(x) \), resulting in the following objective function ~\cite{sharkey2022TakingFeaturesOut, rajamanoharan2024improving}:
\(
L(x) = \|x - \hat{x}\|_2^2 + \lambda \|h(x)\|_1
\),
where \( \lambda \) controls the trade-off between reconstruction fidelity and sparsity. The \( L_1 \) penalty encourages most components of \( h(x) \) to be close to zero, effectively forcing the autoencoder to represent inputs as a sparse linear combination of a small number of feature directions. 
This property makes SAEs particularly useful for high-dimensional data, where they can uncover disentangled, interpretable features.
In the context of combining SAEs with transformers, the input \( x \) for SAE to reconstruct is typically an embedding extracted either from the transformer's residual stream or sometimes from MLP or attention layers~\cite{kissaneInterpretingAttentionLayer2024}.

\section{SAE for Recommendations}

Sequential recommendations are a natural application of sparse autoencoders, as the formulation of the sequential recommendations task is usually very similar to the language modeling task. In recommendations, there is a set of item sequences for each user instead of token sequences in the text data, and the goal is to predict the next item instead of the next token.

\subsection{Training}

\begin{figure*}[ht]
\resizebox{0.85\linewidth}{!}{
\includegraphics[width=\textwidth, trim={0 0.2cm 0 0.5cm},clip]{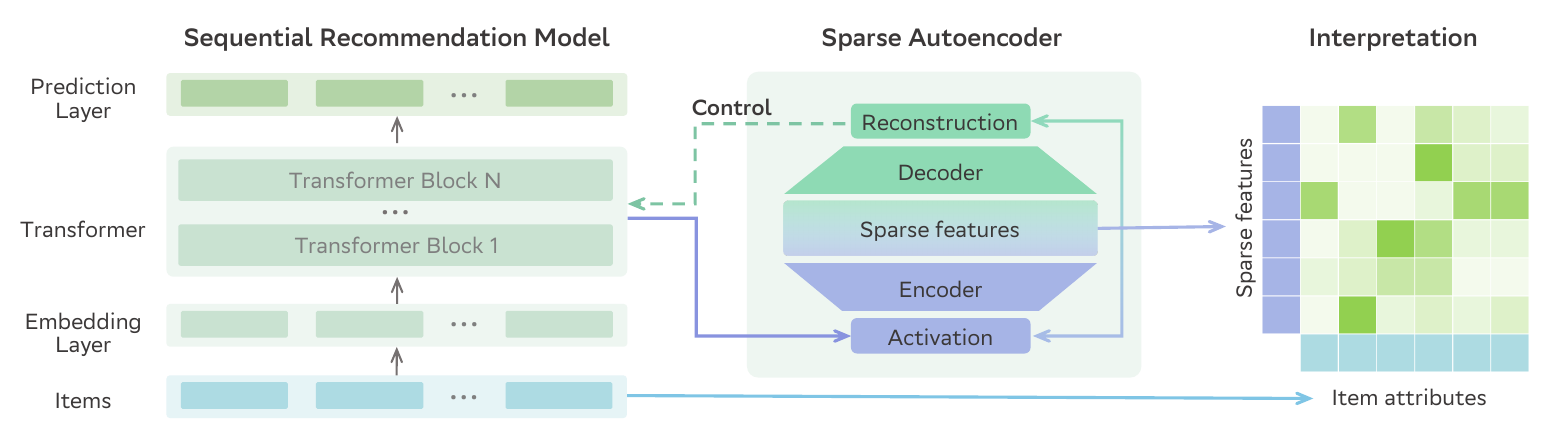}
}
\caption{The schema of SAE for sequential recommendations.}
\label{fig:sae_schema}
\end{figure*}

Our approach for applying SAE to sequential recommendations is illustrated in Figure ~\ref{fig:sae_schema}. Firstly, we train a transformer-based model from scratch on the given dataset. The model is trained solely on user-item interactions without incorporating textual or any other side information. This step differs from the NLP domain, where researchers typically analyze a large model pretrained on a huge amount of data. Secondly, we run part of the data through the model and extract activations from some intermediate transformer layer. The activations from each sequence step are treated as separate data points for training the autoencoder. Next, we train SAE described in Section~\ref{sec:sparse_autoencoders} on these activations.

\subsection{Measuring SAE interpretability}
\label{sec:measuring_interpretability}

The absence of appropriate metrics for interpretability evaluation is one of the main open problems with sparse autoencoders~\cite{makelovPrincipledEvaluationsSparse2024}.
In recommendation systems, predefined attributes are often associated with items, such as movie or music genres or other categorical classifications. These attributes can be used to identify interpretable SAE features by studying the distribution of item classes for the corresponding SAE feature activations. We expect that a well-working SAE has learned features that at least partially align with these predefined attributes. The strength of this alignment serves as an indicator of the interpretability of the autoencoder. Based on this, we introduce metrics to quantify and evaluate the overall interpretability of the trained SAE.

We compute the following metrics for each pair of SAE features and item attributes:
\begin{itemize}
    \item \textit{Pearson correlation coefficient} between binary attribute and continuous feature activation value (this special case of Pearson correlation is also called point-biserial correlation~\cite{tate1954correlation}). The correlation is a good indicator in this case, as we do not expect any complex nonlinear dependencies between the attributes and the feature activations. We are looking for a simple case where the higher the activation, the higher the probability of the attribute.
    \item \textit{ROC AUC}. We consider the attribute value as a target variable, the feature activation as a prediction, and compute the area under the ROC curve.
    \item \textit{Sensitivity} (we define it similarly to~\cite{bricken2023towards}) - a fraction of examples with this attribute where the feature is active. High sensitivity indicates that the feature captures the attribute in its entirety rather than focusing on its specific components.
\end{itemize}

According to our observations, correlation is more aligned with the quality of correspondence between the feature and the attribute than ROC AUC. ROC AUC is not sensitive to class imbalance and may favor features with low precision.

We sort a matrix of these metrics for each feature-attribute pair by the metric value for a given attribute, thereby identifying the top features for each attribute. Examining the metric values for a given feature across all other attributes allows to evaluate the feature's monosemanticity. A comparison with the metric values for the top neurons of the original transformer layer demonstrates that SAE features significantly enhance interpretability. Moreover, we calculate the mean metric value across the top features for all attributes, using it as an overall measure of the interpretability of the SAE.

\subsection{Controlling the Model Behavior}

Although the model interpretation itself is an interesting research question, the ability to control the model behavior is of greater importance for practical use. If we find a feature corresponding to a given item attribute or another desired property, then we can change the activation of this feature. If this feature also has a causal downstream effect, an increase in the activation value will lead to a greater presence of this attribute in the model predictions. Conversely, a decrease in the activation value will make this attribute less present in the recommendations.

This way it is possible to provide recommendation services with an ``equalizer'' to adjust the presence of specific attributes and make more personalized recommendations. For example, it is possible to change the presence of a particular music genre or fill the music stream with more relaxing or more energetic music depending on the user's mood and context, or we can mitigate popularity bias and increase the diversity of recommendations by decreasing the proportion of the most popular genre.

We perform intervention in the forward pass of the model during inference in the following way. We take the activations of the selected transformer layer, encode them with the SAE encoder, set the activation of the selected SAE feature to some desired value, reconstruct the original layer with the SAE decoder, and replace the layer with this reconstruction. This approach is commonly referred to as steering~\cite{bricken2023towards}. If the SAE feature corresponds to some attribute, the model outputs will change accordingly.

\section{Experimental Setting}
\label{sec:experimental_settings}

Experiments are conducted on the Movielens-20M~\cite{harper2015movielens} and Music4All~\cite{santana2020music4all} datasets. The movie and music domains are well-suited for studying model interpretability and control, as user preferences over content categories like genres are often meaningful and diverse. The ability to guide recommendations in such settings can be valuable for improving user experience and personalization. Movielens-20M contains user ratings of movies, as well as their attributes: title, genres, and release year. We additionally include the movie’s language in our analysis. Music4All contains user listening history along with various song attributes, including genres. The genre taxonomy is rather messy, as there are 853 unique values. So, we take 20 most popular genres in the analysis.

We split the datasets into three parts. The train-test split is performed both by time and by users, so the test set contains 10\% of the most recent users. This part is used only for the evaluation of SAE features and the underlying recommendation model. All previous interactions are taken into the training dataset, which is further split into two parts by users. We use 60\% of users for training a recommendation model and 40\% of users for training a sparse autoencoder on activations of this model.

We use two transformer-based models: GPTRec~\cite{petrov2023generative}, which follows the transformer decoder architecture and is very similar to the widely-used SASRec~\cite{kang2018self}, and BERT4Rec~\cite{sun2019bert4rec}, which is based on the transformer encoder architecture. This allows us to cover both major classes of transformer-based models used in sequential recommendations. We set the number of transformer layers to 3, the number of attention heads to 2, and the hidden size to 64.

SAE is trained on activations from the output of the first transformer block, as this layer has slightly better interpretability metrics. Nevertheless, selecting other layers leads to similar results. We normalize activations by subtracting the mean and dividing by the standard deviation. Normalization is a common practice~\cite{gaoScalingEvaluatingSparse2024,rajamanoharan2024jumping}, which makes it easier to transfer hyperparameters and compare different runs on different layers. SAE is trained with Adam optimizer, and the learning rate is 1e-3. We tried different SAE dictionary size values ranging from 64 to 8196 and different weights of L1 penalty ranging from 0.01 to 1. The choice of these hyperparameters is discussed in the following sections.

Since the results are consistent across models and datasets, we report results for GPTRec on the MovieLens-20M dataset by default, unless stated otherwise. The appendix contains additional results for BERT4Rec and the Music4All dataset.
We use \textit{dictionary\_learning} repository\footnote{\url{https://github.com/saprmarks/dictionary_learning}} for SAE training, \textit{NNsight} library~\cite{fiottokaufman2024nnsightndifdemocratizingaccess} for model intervention, and \textit{RePlay} library~\cite{vasilev2024replay} for preprocessing and metrics calculation.

\section{SAE Reconstruction Properties}

A sparse autoencoder is designed to reconstruct the activations of a transformer layer, but perfect reconstruction is not the main goal -- the key objective is to learn interpretable features. If reconstruction quality is too low,  SAE may fail to capture meaningful properties of the original model. At the same time, sparsity plays a crucial role: an overly sparse model may lack the capacity for accurate reconstruction, while too many active features can make the representation trivial and reduce interpretability. Therefore, effective SAE design requires balancing reconstruction quality and sparsity.

To determine whether SAE learns the features relevant to the downstream recommendation task, we replace the activations of the transformer layer with their reconstructed values during the forward pass and look at how this operation affects the model predictions. In particular, we measure the change in recommendation metrics for the next-item prediction task. We consider the last interaction of each user in the test set (see Section~\ref{sec:experimental_settings}) as ground truth value and compute metrics on these interactions.

\begin{table}[t!]
\caption{SAE metrics for different values of L1 regularization: RMSE, a fraction of explained variance, L0 (sparsity, average number of active features per example), and recommendation metrics for the model with one layer output replaced by its reconstruction. The first row corresponds to the
original model without reconstruction.}
\label{tab:reconstruction_metrics_l1}
\centering
\small
\resizebox{\columnwidth}{!}{
\begin{tabular}{@{}lrlrrrr@{}}
\toprule
\multicolumn{1}{c}{\textbf{L1}} & \multirow{2}{*}{\textbf{RMSE}} & \textbf{Explained} & \multicolumn{1}{c}{\multirow{2}{*}{\textbf{L0}}} & \multicolumn{3}{c}{\textbf{Recommend. metrics @10}} \\
\cmidrule{5-7}
\multicolumn{1}{c}{\textbf{weight}}&&\multicolumn{1}{c}{\textbf{variance}} & &\textbf{NDCG}&\textbf{HitRate}&\textbf{Coverage} \\
\midrule
Original & -- & -- & -- & 0.100 & 0.174 & 0.217 \\
0.001 & 0.020 & 1.000 & 821.5 & 0.100 & 0.174 & 0.217 \\
0.05 & 0.008 & 1.000 & 137.3 & 0.100 & 0.174 & 0.217 \\
0.1 & 0.178 & 0.958 & 102.1 & 0.098 & 0.172 & 0.210 \\
0.2 & 0.340 & 0.850 & 29.0 & 0.095 & 0.168 & 0.190 \\
0.3 & 0.501 & 0.676 & 9.3 & 0.087 & 0.156 & 0.161 \\
0.4 & 0.724 & 0.337 & 2.7 & 0.058 & 0.112 & 0.114 \\
0.5 & 0.880 & 0.060 & 0.6 & 0.041 & 0.083 & 0.104 \\
1.0 & 0.922 & 0.0001 & 0.5 & 0.039 & 0.079 & 0.103 \\
\bottomrule
\end{tabular}
}
\end{table}

Table~\ref{tab:reconstruction_metrics_l1} contains root mean squared error (RMSE), a fraction of explained variance, an average number of active features (sparsity), as well as NDCG@10, HitRate@10, and Coverage@10 for different values of L1 regularization and dictionary size of 2048.
L1 regularization is the most influential hyperparameter, as small changes in its values have a strong effect on all metrics. For further experiments we set L1 regularization to 0.1, as for this value the reconstruction is good enough, but not perfect. Figure~\ref{fig:corr_dependency} from Section~\ref{sec:overall_metrics} confirms that for this value, interpretability is the best.

As for dictionary size, we found that most of the results are not very sensitive to it for the fixed value of L1 penalty. Reconstruction error remains stable for dictionary sizes from 512 to 8192, while the number of active features is stable for sizes from 1024 to 8192 and becomes smaller for smaller dictionary sizes. We choose a dictionary size of 2048 for further experiments.

\section{Interpretation}
\label{sec:interpretation}

We focus our interpretability metrics on genres, as they are clear, well-understood, and widely used in practice. Enabling control over genre-level attributes can be particularly relevant for real-world applications. Additionally, we provide interpretability evaluation using other item attributes in Appendix~\ref{sec:appendix_attributes}.

\subsection{Results}

\begin{figure}[h!]
    \centering
    \includegraphics[width=\linewidth]{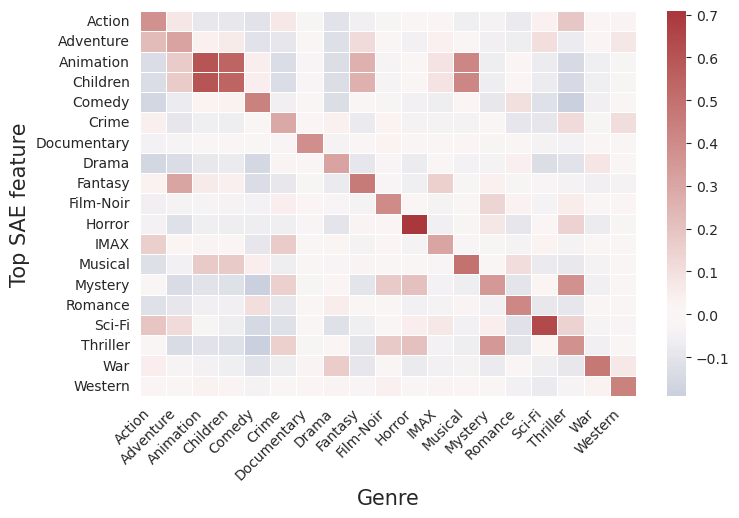}
    \caption{Correlation between genres and top feature (the feature with maximum correlation) for each genre. One row corresponds to one feature and contains its correlations with all genres.}
    \label{fig:genre_corr_heatmap}
\end{figure}

\subsubsection{Finding and analyzing top features for genres.}
\label{sec:finding_top_neurons}

We follow the approach described in Section~\ref{sec:measuring_interpretability} and compute correlation, ROC AUC, and sensitivity between each SAE feature and genre. For each genre, we identify the top feature with the highest correlation. Figure~\ref{fig:genre_corr_heatmap} shows correlations of these top features with all genres on the Movielens-20M dataset. The features are mostly monosemantic and correlate with one or two similar genres. Animation and Children are the most interconnected genres since they both share the same top feature, which is expected as animated movies are often intended for children. The same connection holds for Thriller and Mystery. Other patterns (e.g., Adventure feature correlates with Action, Fantasy with Adventure, War with Drama) are also consistent with genre semantics.

\begin{table}[t!]
\caption{Metrics for SAE features with maximum correlation for each genre and genre popularity (proportion of data points with given genre). The data is sorted by correlation. The last row shows the averages across all genres, which are considered as overall interpretability metrics.}
\label{tab:metrics_by_genre}
\centering
\small
\resizebox{\columnwidth}{!}{
\begin{tabular}{@{}lrrrr@{}}
\toprule
\textbf{Genre}       & \textbf{Correlation} & \textbf{ROC AUC} & \textbf{Sensitivity} & \textbf{Genre popularity} \\ 
\midrule
Horror & 0.708 & 0.919 & 0.885 & 0.066 \\
Sci-Fi & 0.637 & 0.909 & 0.868 & 0.182 \\
Animation & 0.598 & 0.910 & 0.901 & 0.083 \\
Children & 0.545 & 0.873 & 0.840 & 0.095 \\
Musical & 0.493 & 0.791 & 0.609 & 0.038 \\
War & 0.476 & 0.819 & 0.745 & 0.061 \\
Fantasy & 0.462 & 0.707 & 0.568 & 0.135 \\
Comedy & 0.437 & 0.744 & 0.550 & 0.331 \\
Western & 0.436 & 0.761 & 0.576 & 0.015 \\
Romance & 0.413 & 0.715 & 0.481 & 0.175 \\
Film-Noir & 0.401 & 0.711 & 0.455 & 0.011 \\
Documentary & 0.383 & 0.697 & 0.391 & 0.011 \\
Thriller & 0.378 & 0.714 & 0.525 & 0.278 \\
Action & 0.378 & 0.686 & 0.437 & 0.311 \\
Mystery & 0.350 & 0.756 & 0.655 & 0.099 \\
Drama & 0.315 & 0.658 & 0.369 & 0.470 \\
Adventure & 0.313 & 0.652 & 0.420 & 0.255 \\
IMAX & 0.310 & 0.597 & 0.235 & 0.061 \\
Crime & 0.290 & 0.637 & 0.343 & 0.195 \\
\midrule
Mean & 0.438 & 0.750 & 0.571 & 0.151 \\
\bottomrule
\end{tabular}
}
\end{table}

Table~\ref{tab:metrics_by_genre} reports interpretability metrics for these top features. The results vary greatly depending on the genre. For some genres, such as Horror and Sci-Fi, the correlation and ROC AUC are high, whereas for others they are significantly lower. For example, broader and more loosely defined genres like Action, Drama, and Adventure have lower correspondence with features, likely because they appear too frequently and across different contexts.

The results for BERT4Rec and the Music4All dataset are similar and provided in the appendix.

\subsubsection{
Overall interpretability metrics.}
\label{sec:overall_metrics}

\begin{figure}[ht!]
 \centering
 \begin{subfigure}{0.49\linewidth}
     \includegraphics[width=\textwidth]{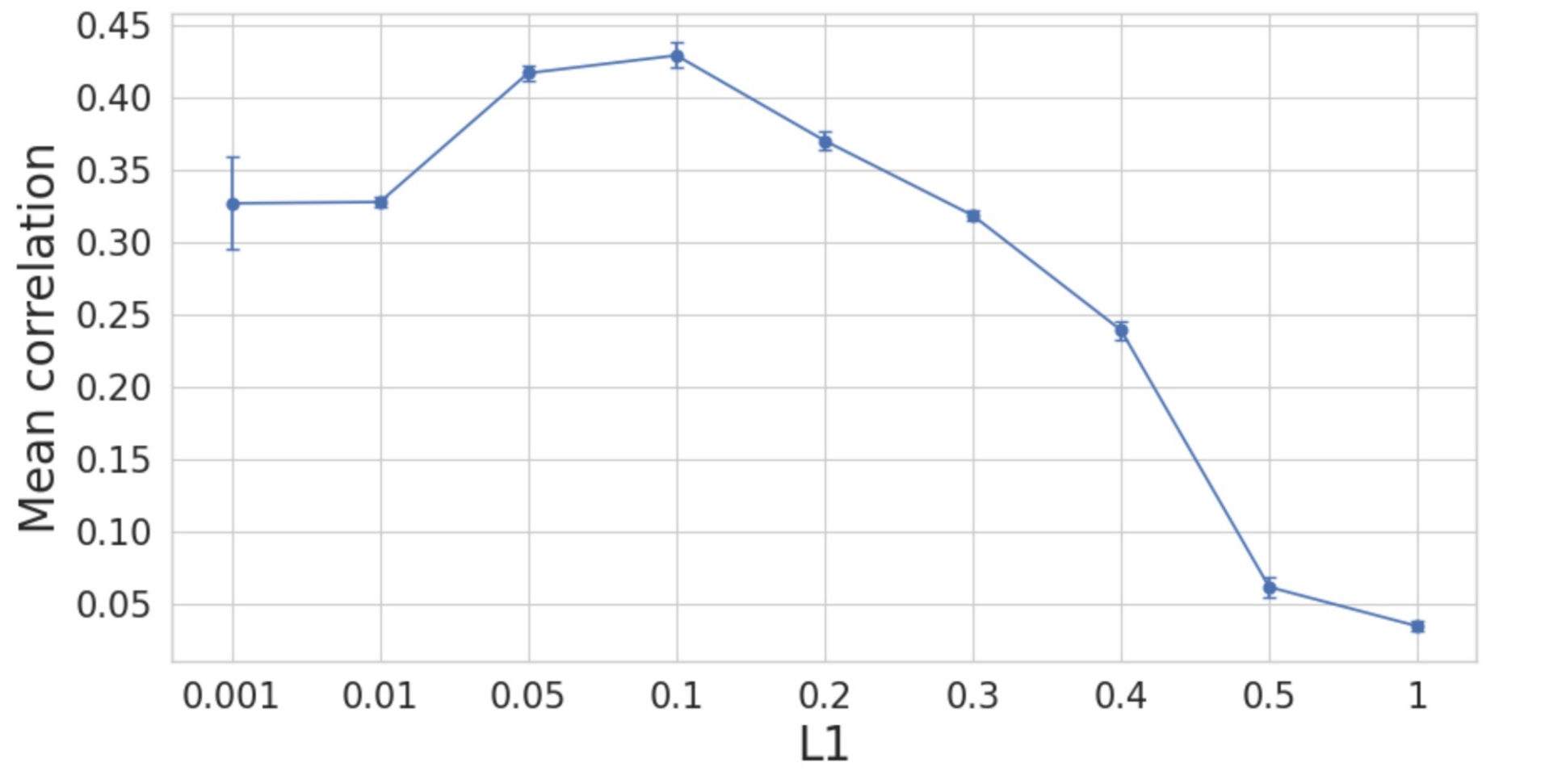}
     \caption{L1 regularization}
 \end{subfigure}
 \begin{subfigure}{0.49\linewidth}
     \includegraphics[width=\textwidth]{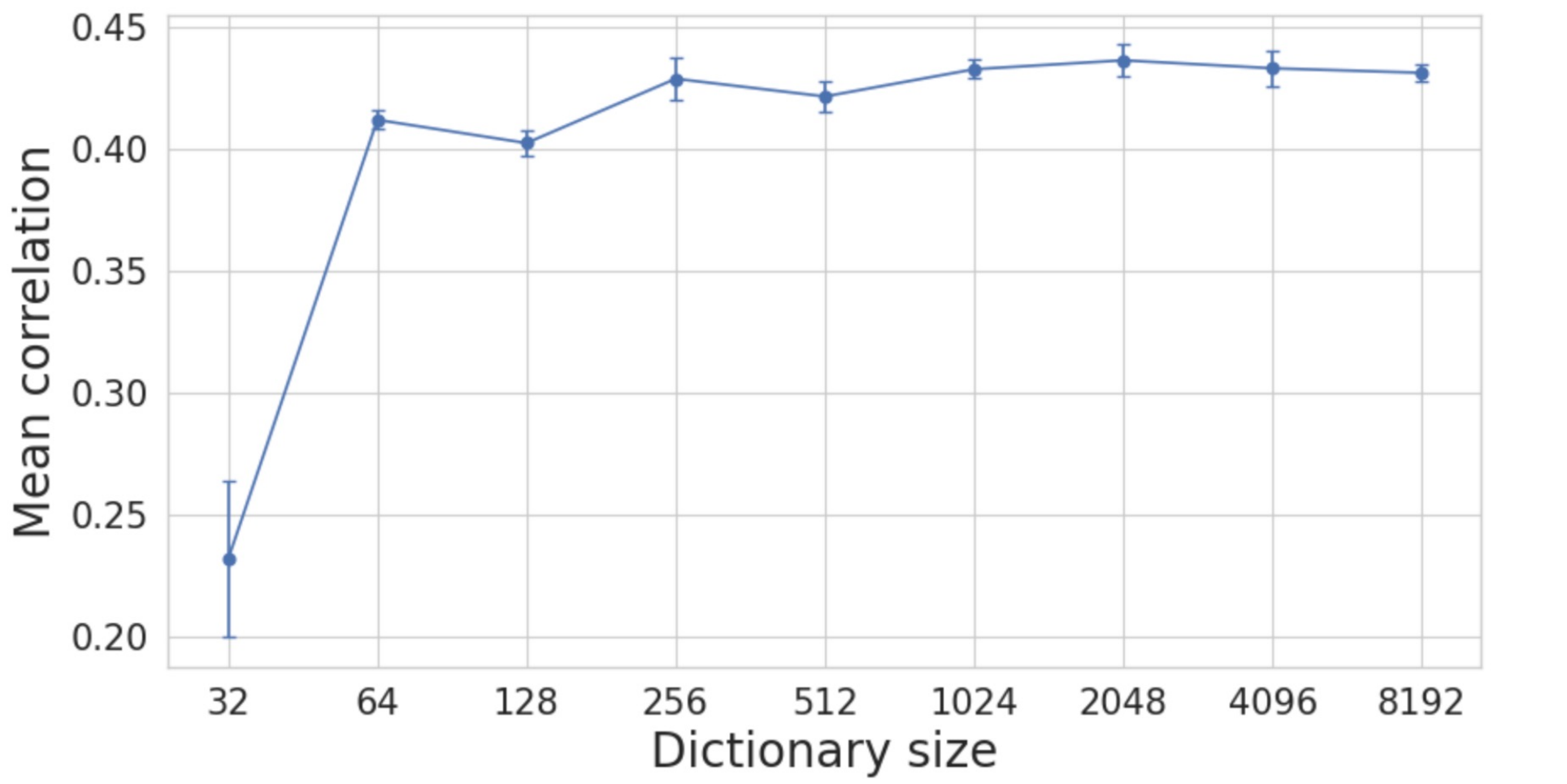}
     \caption{Dictionary size}
 \end{subfigure}
 \caption{Dependency of the SAE interpretability metric (mean correlation) on the SAE parameters. For the L1 plot, the dictionary size is set to 2048, and for the dictionary size plot, the L1 is set to 0.1.}
 \label{fig:corr_dependency}
\end{figure}

The last line of Table~\ref{tab:metrics_by_genre} contains mean metrics across all genres.
These values reflect SAE interpretability and can be used as a guideline for selecting SAE hyperparameters. Figure~\ref{fig:corr_dependency} illustrates the dependency of mean correlation on L1 penalty and SAE dictionary size. We make 5 runs for each point and average them, as we found that the mean correlation can fluctuate across runs. The mean correlation is lower for very low L1 values with perfect reconstruction quality and for high L1 values with bad reconstruction quality (see Table~\ref{tab:reconstruction_metrics_l1}). For a dictionary size above 256, the mean correlation is stable.

\subsubsection{Comparison with the neurons of the original transformer layer.}
\label{sec:original_vs_sae}

\begin{figure}[ht!]
 \centering
 \begin{subfigure}{\linewidth}
     \includegraphics[width=\textwidth]{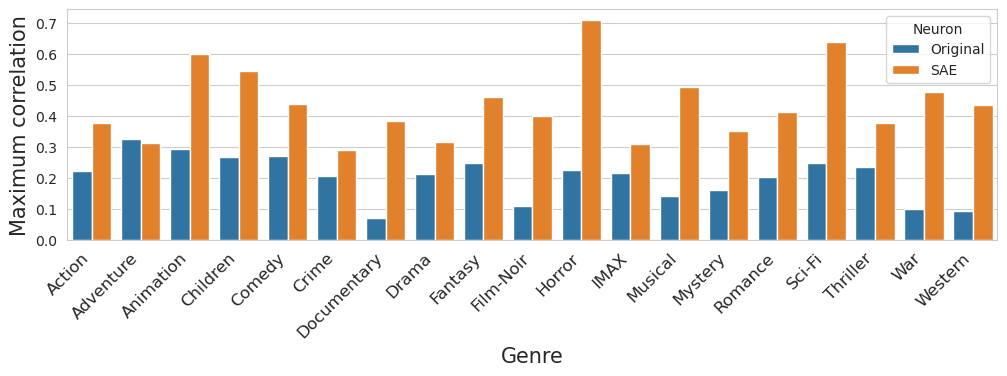}
     \caption{Movielens-20M}
     \label{fig:corr_original_vs_sae_movielens}
 \end{subfigure}
 \begin{subfigure}{\linewidth}
     \includegraphics[width=\textwidth]{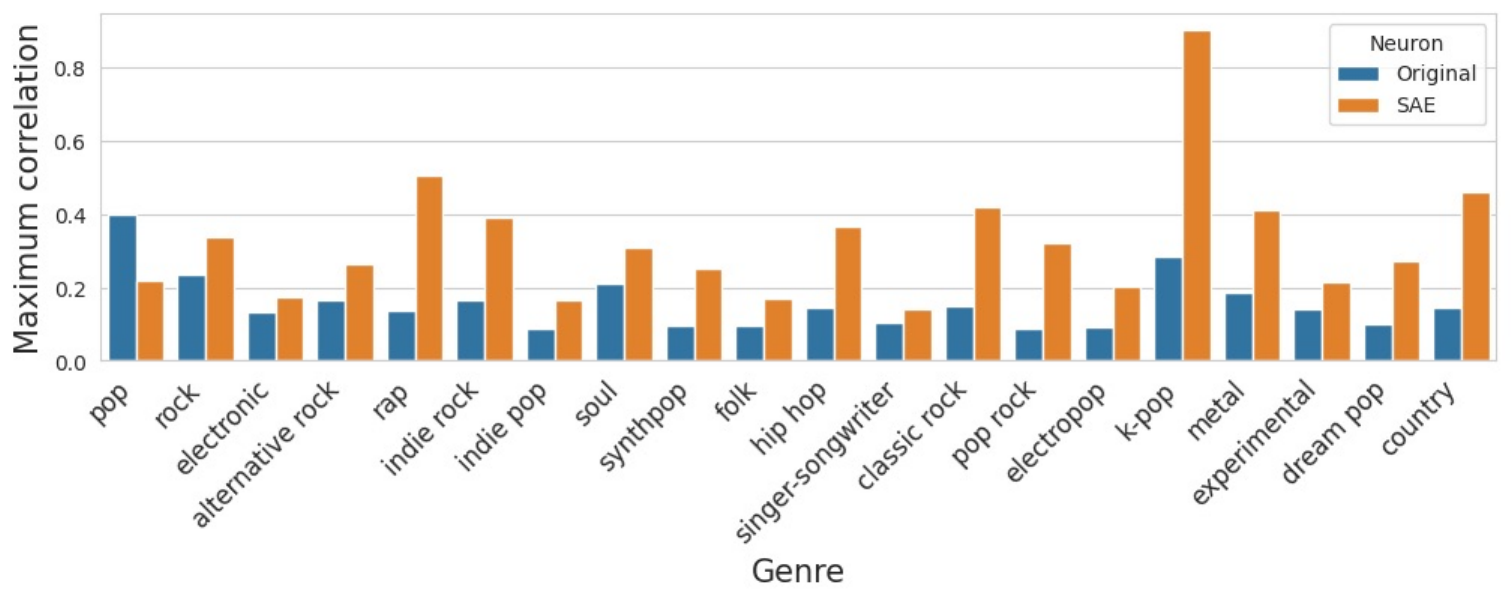}
     \caption{Music4all}
     \label{fig:corr_original_vs_sae_music4all}
 \end{subfigure}
 \caption{Comparison between interpretability of SAE features and neurons of the original transformer layer. The maximum correlation for each genre is blue for the transformer layer and orange for SAE. For the Music4all dataset, the genres are sorted by their popularity in the dataset.}
 \label{fig:corr_original_vs_sae}
\end{figure}

It is important to check whether SAE features are more interpretable than the neurons of the original transformer layer on which SAE was trained. In order to do this, we perform the same procedure, compute the correlation between neurons of the transformer layer and genres, and find the top neurons with the maximum correlation for each genre.

Figure~\ref{fig:corr_original_vs_sae_movielens} contains these correlation values for top neurons of the transformer layer along with top SAE features for the Movielens-20M dataset. The association with the SAE feature is much higher for all genres except Adventure. Correlation with neurons of the transformer layer is especially low for rare genres, e.g. Documentary, Film-Noir, and Western, while SAE is able to learn good features for these cases.

\begin{table}[t!]
\caption{Comparison of SAE features with original transformer layer neurons. Minimum, mean, and maximum correlation values for top features/neurons.}
\label{tab:sae_vs_transformer}
\centering
\small
\resizebox{\columnwidth}{!}{
\begin{tabular}{@{}llrrrrrr@{}}
\toprule
\multirow{2}{*}{\textbf{Dataset}} & \multirow{2}{*}{\textbf{Model}} & \multicolumn{3}{c}{\textbf{Transformer Layer}} & \multicolumn{3}{c}{\textbf{SAE Features}} \\
\cmidrule(lr){3-5} \cmidrule(lr){6-8}
& & \textbf{min} & \textbf{mean} & \textbf{max} & \textbf{min} & \textbf{mean} & \textbf{max} \\ 
\midrule
Movielens-20M & GPTRec  & 0.07 & 0.20 & 0.33 & 0.29 & 0.44 & 0.71 \\
Movielens-20M & BERT4Rec & 0.04 & 0.14 & 0.22 & 0.16 & 0.29 & 0.53 \\
Music4all & GPTRec  & 0.09 & 0.16 & 0.40 & 0.14 & 0.32 & 0.90 \\
\bottomrule
\end{tabular}
}
\end{table}

Figure~\ref{fig:corr_original_vs_sae_music4all} contains the correlation values for the Music4all dataset. Association with SAE features is significantly higher for most of the genres, with some genres being more successful than others. For pop music, the correlation with the SAE feature is low, while the correlation with the original neuron is much higher. This is because it is the most popular genre, which is reflected in the model, whereas for SAE it is too broad, so the features capture only its subsets.

The corresponding figure for BERT4Rec is provided in Appendix~\ref{sec:appendix_bert4rec}. Table~\ref{tab:sae_vs_transformer} summarizes the results, showing the minimum, mean, and maximum correlation for both the neurons of the transformer layer and the SAE features. Although the absolute values for SAE features for BERT4Rec are lower than those for GPTRec, this is due to weaker genre alignment in the original BERT4Rec layer. Nevertheless, SAE still clearly improves interpretability.

\subsection{Case Studies}

\begin{figure}[ht!]
 \centering
 \begin{subfigure}{0.4\columnwidth}     \includegraphics[width=\textwidth]{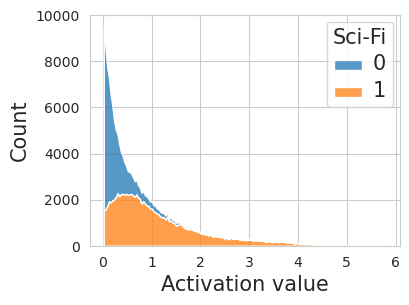}
     \caption{Sci-Fi}
 \end{subfigure}
 \begin{subfigure}{0.4\columnwidth}
     \includegraphics[width=\textwidth]{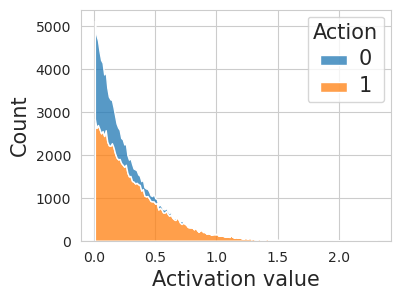}
     \caption{Action}
 \end{subfigure}
 \caption{Distribution of the activation values for SAE features corresponding to a given genre. The orange color corresponds to cases when the genre is present; the blue color corresponds to cases when the genre is absent.}
 \label{fig:activations}
\end{figure}

\subsubsection{Features corresponding to a given genre.}
\label{sec:case_studies_genre}

\begin{figure*}[ht!]
    \centering
        \hfill
        \begin{subfigure}{0.16\textwidth}
            \includegraphics[width=\textwidth, height=6.5cm, keepaspectratio]{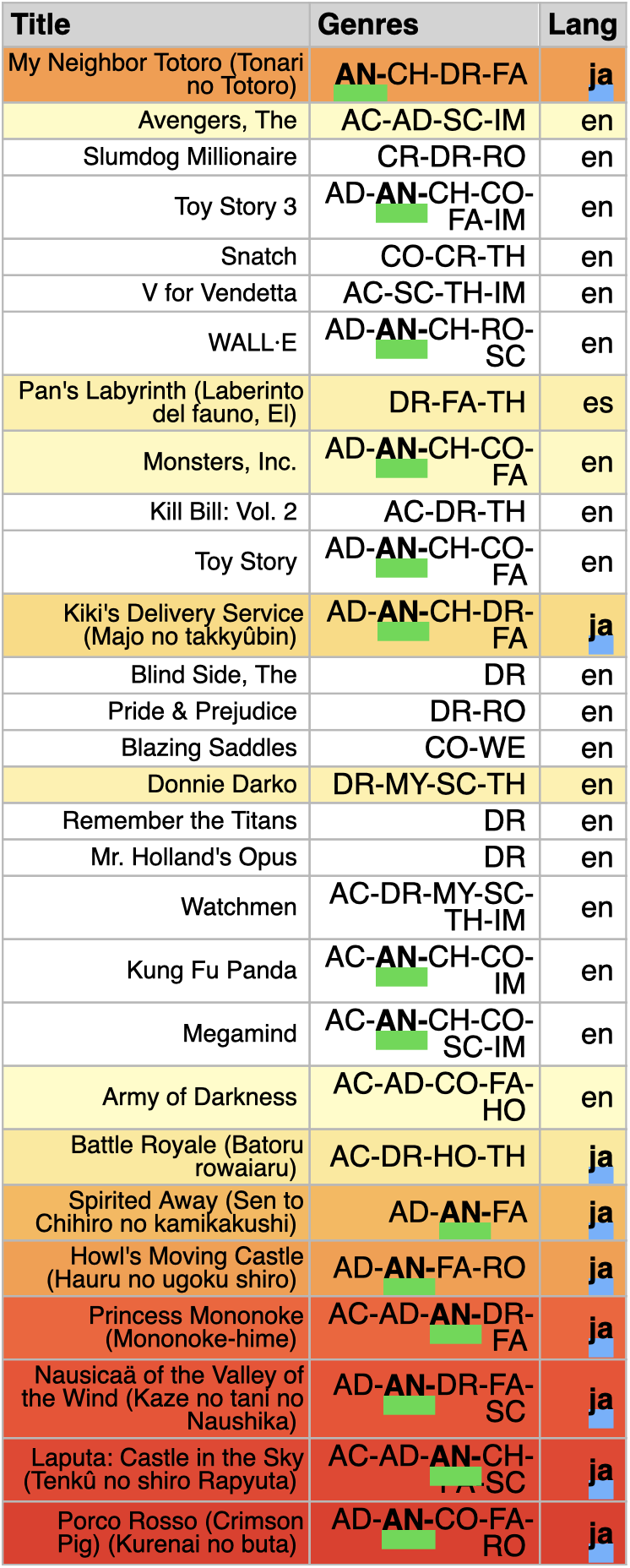}
            \caption{Anime}
            \label{fig:case_anime}
        \end{subfigure}
        \hfill
        \begin{subfigure}{0.1\textwidth}
            \includegraphics[width=\textwidth, height=6.5cm, keepaspectratio]{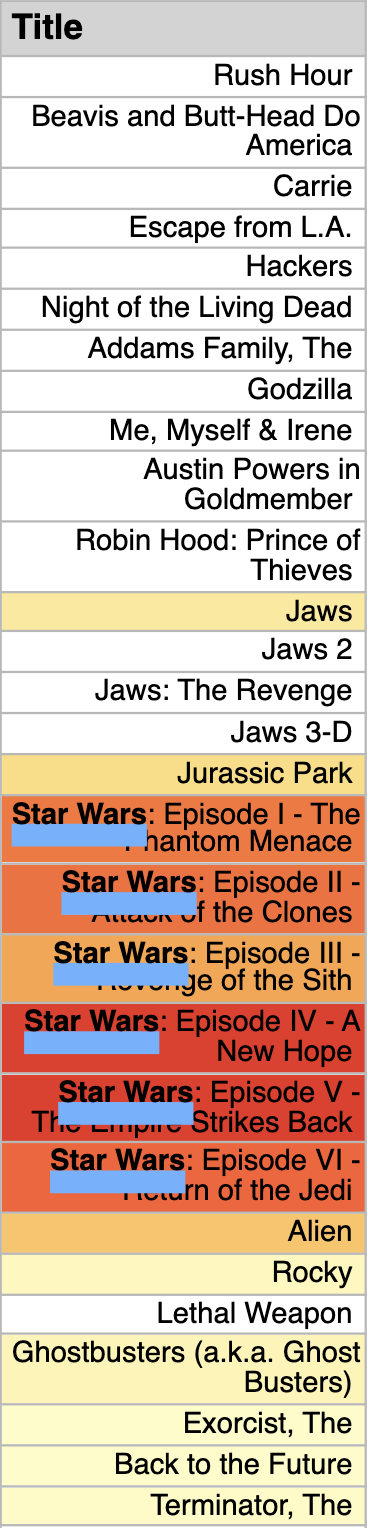}
            \caption{Star Wars}
            \label{fig:case_starwars2}
        \end{subfigure}
        \hfill
        \begin{subfigure}{0.22\textwidth}
            \includegraphics[width=\textwidth, height=6.5cm, keepaspectratio]{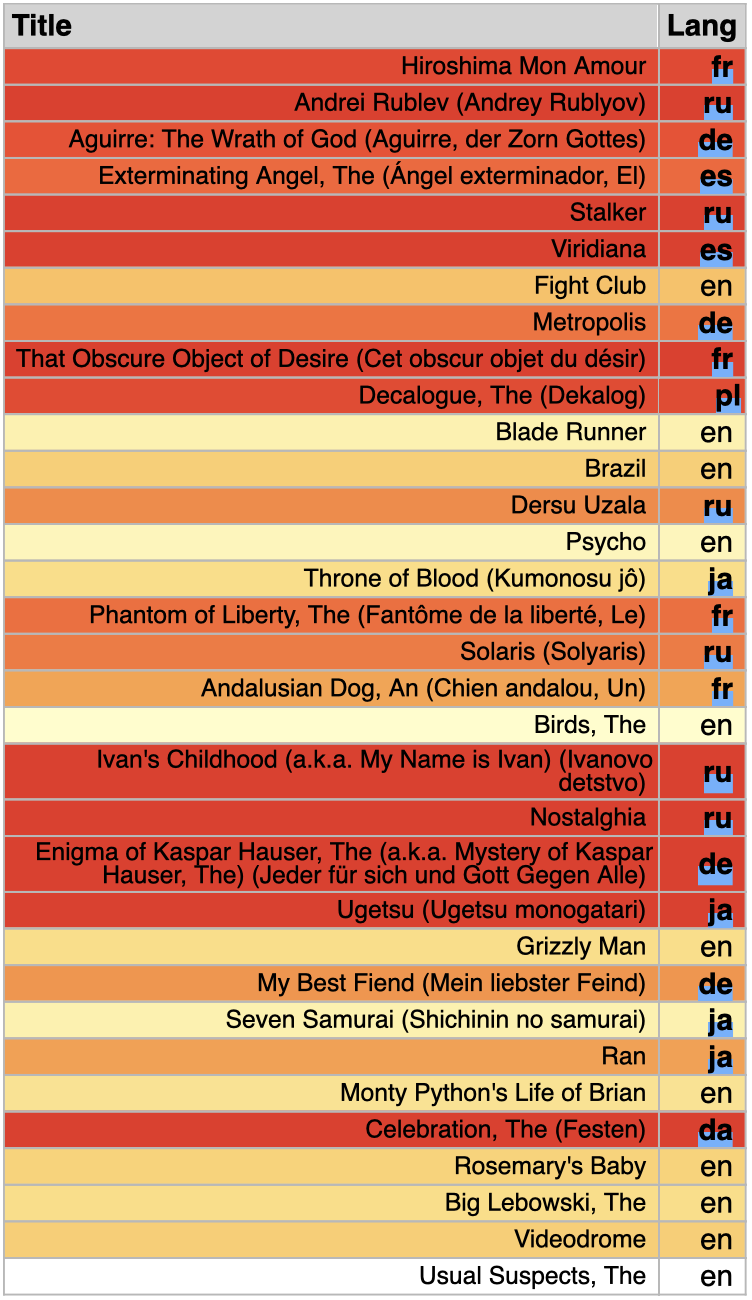}
            \caption{Non-English}
            \label{fig:case_foreign}
        \end{subfigure}
        \hfill
        \begin{subfigure}{0.16\textwidth}
            \includegraphics[width=\textwidth, height=6.5cm, keepaspectratio]{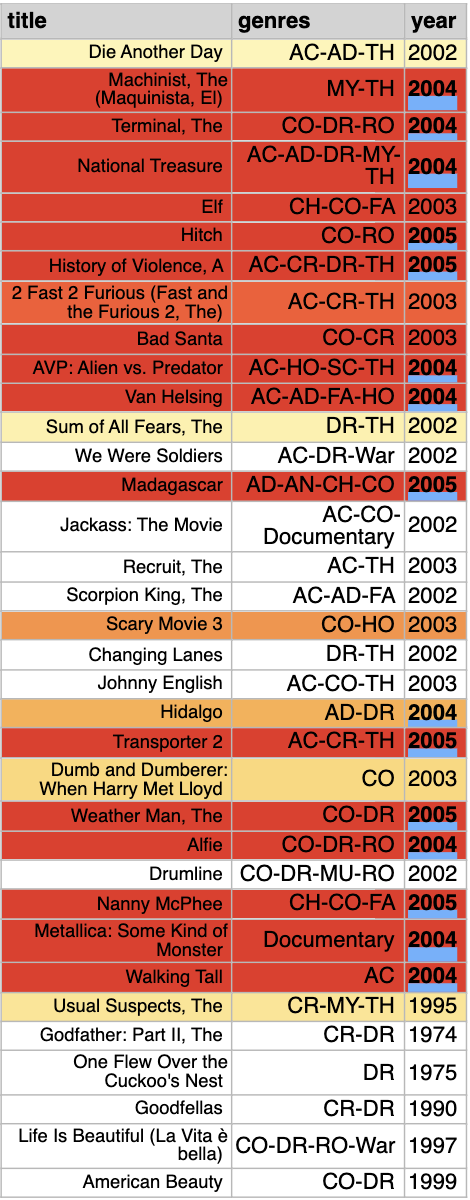}
            \caption{Movies of 2004-2005}
            \label{fig:case_2004}
        \end{subfigure}
        \hfill
        \begin{subfigure}{0.2\textwidth}
            \includegraphics[width=\textwidth, height=6.5cm, keepaspectratio,trim={0 8.12cm 0 0},clip]{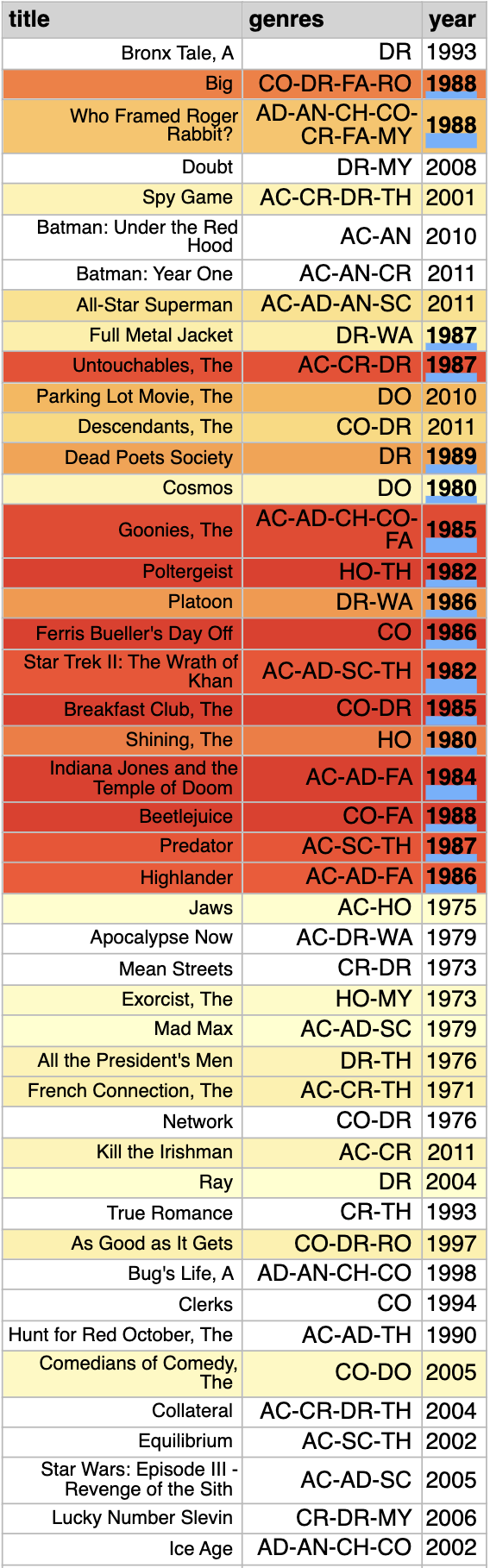}
            \caption{80's movies}
            \label{fig:case_80s}
        \end{subfigure}
        \hfill
        \caption{Examples of SAE features. Each subplot represents a user interaction sequence\protect\footnotemark[1]. Yellow-to-red background color shading illustrates the degree of a feature activation on a given item (red is the strongest). (a) A Japanese Anime. Green highlights animation genre, blue indicates the Japanese language of origin. (b) Star Wars series. (c) non-English language movies. (d) and (e) features for movies released around 2004-2005 and 1980s, respectively.}
\label{fig:case_samples}    
\end{figure*}

At first, we examine the top features found in section~\ref{sec:finding_top_neurons}. Figure~\ref{fig:activations} presents histograms of positive per-item feature activations for features exhibiting the highest correlation with Sci-Fi and Action. Histogram colors indicate whether the activation happened on an item of the associated genre. The proportion of such items increases notably with higher activation levels, which is reminiscent of the behavior of interpretable SAE features observed in NLP models~\cite{cunninghamSparseAutoencodersFind2023a,bricken2023towards}.
We found the same pattern for all other genres. Activation of a given feature above some threshold typically indicates the presence of respective genre. However, the threshold varies between features, as well as the scale of activation values.

Interestingly, the proportion of the Action genre is high despite its relatively low correlation with the SAE feature, suggesting that the feature captures only a subset of this genre -- likely due to the genre's wide-ranging diversity and scope.

\subsubsection{Diverse interpretable features.}

To investigate further, we analyzed various SAE features in depth. For each feature we summed activations of the top 100 samples with the strongest responses and normalized by all activations in each sample, yielding the activation probability $p_{top100}$. We found that activation patterns of a substantial proportion of features can be interpreted in a meaningful way. Remarkably, these features were learned by SAE in an unsupervised manner.

Although the previous features correspond to the entire genre classes in the dataset, some features capture categorizations that go beyond these predefined classes. For example, Figure~\ref{fig:case_anime} illustrates the activation pattern of a feature that activates on Japanese Anime items but remains largely inactive on Animation movies in other languages. The distinction is demonstrated in Figure~\ref{fig:case_anime} by highlighting the Animation genre in green and the Japanese original language in blue. Items with the Animation attribute account for $p_{top100} = 0.633$ of all activations, while items labeled with Japanese as the original language comprise $p_{top100} = 0.582$.

Another interesting example is the feature that corresponds to the "Star Wars" movie series. In Figure~\ref{fig:case_starwars2}, "Star Wars" movies are highlighted in blue across two representative sequences: one where the feature is active on movies appearing consecutively, and another where the feature activates on a single "Star Wars" movie within its context. This feature's activation proportion for the items containing "Star Wars" in title is $p_{top100} = 0.693$. There are other features with even higher such $p_{top100}$ for the "Star Wars", but they only respond to the specific movie from the series (e.g., features that activate on "Episode I" with $p_{top100} = 0.966$ and on "Episode IV" with $p_{top100} = 0.978$ have no overlap, $p_{top100} = 0$ for each other's title), rather than finding the common direction as in the case of the "Star Wars" series feature. In general, we found a number of features that activate predominantly on various specific (usually popular) movies.

There are multiple other curious representations learned by SAE. For instance, Figure~\ref{fig:case_foreign} shows a feature that activates on non-English language movies. Other attributes of items, such as movie release year, were also reflected in SAE features. Figures~\ref{fig:case_2004} and~\subref{fig:case_80s} show two such features: the former activates on movies released around 2004 -- 2005, while the latter captures movies from the 1980s, both in an apparently genre-agnostic manner.

\section{Control}

In this section, we determine how the modification of SAE activation values influences the model predictions.

\subsection{Examples}

\begin{figure}[ht!]
\centering
\begin{subfigure}{\linewidth}
\includegraphics[width=\linewidth]{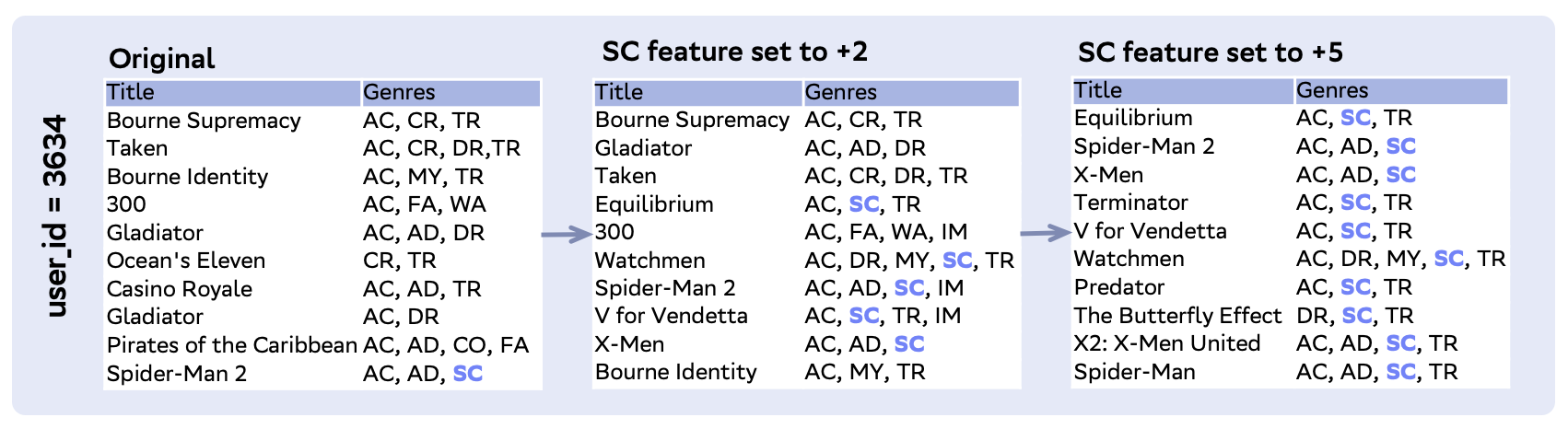}
\end{subfigure}
\begin{subfigure}{\linewidth}
\includegraphics[width=\linewidth]{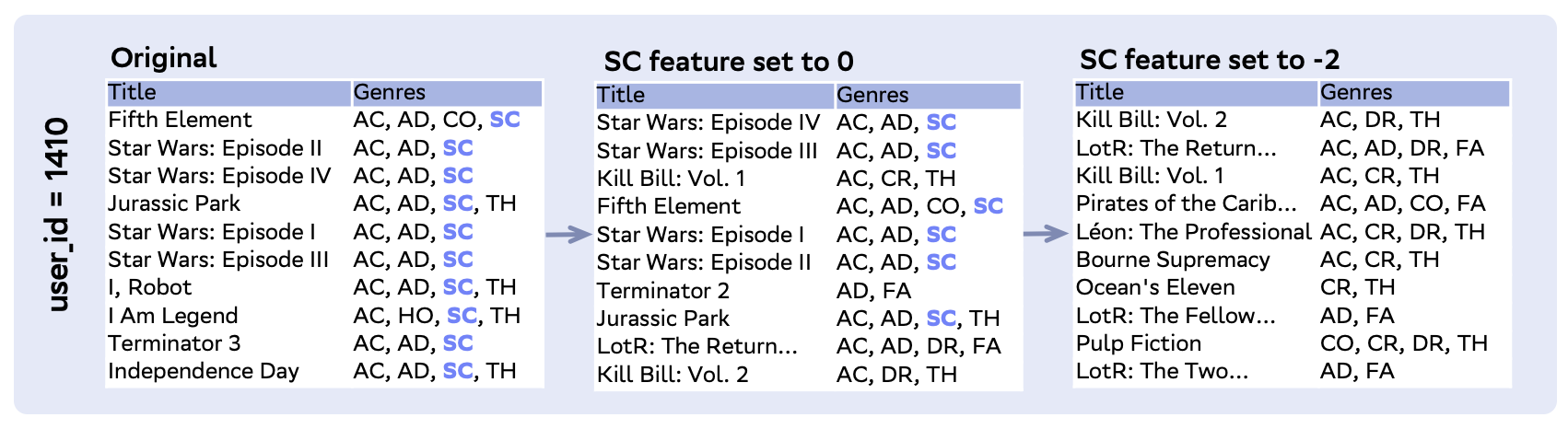}
\end{subfigure}

\caption{Examples of changes in the recommendations with Sci-Fi (SC) feature\protect\footnotemark[1]. Recommendation lists before and after changing the feature activation.}
\label{fig:intervention_examples}
\end{figure}

Firstly, we show several examples of successful modification of the model behavior. We demonstrate how recommendations for a given user are influenced by changes in the activation of the SAE feature from Section~\ref{sec:case_studies_genre}, which corresponds to the Sci-Fi genre. Figure~\ref{fig:intervention_examples} contains recommendation lists obtained before and after the intervention.

For user 3634, Action and Thriller genres predominate in the recommendations. After we set the activation of the Sci-Fi feature to +2, there are several movies from the original list and several new Sci-Fi movies. After we set the activation of the Sci-Fi feature to +5 (this is a very high value, exceeding most of the real examples in the dataset), all movies in the recommendation list belong to Sci-Fi. Interestingly, Action and Thriller remain among the most represented genres after the intervention. This means that the intervention performs a kind of soft control and shifts predictions toward Sci-Fi while taking into account other user interests. Thus this approach opens up an opportunity to mix the desired attributes to varying degrees. 

It is also possible to shift the recommendations in the opposite direction and remove the selected genre. The original recommendation list for user 1410 consists of Sci-Fi movies. If we set the activation value of the Sci-Fi feature to 0, part of Sci-Fi movies disappears from the recommendations. When we set the activation value of the Sci-Fi feature to -2, all Sci-Fi movies will disappear from the recommendations. However, Action, Adventure, and Thriller, which were present in addition to Sci-Fi, remain after intervention again. It is worth noting that all SAE activations are non-negative. However, to turn off the desired feature, we can set the activation to a large negative value. This way, it is possible to shift along the corresponding direction far enough to achieve the expected effect.

These examples show promising results, but it is necessary to check whether these results generalize to all users. We perform a comprehensive evaluation in the next section.

\footnotetext{Genre abbreviations: 
Action (AC), Adventure (AD), Animation (AN), Children (CH), Comedy (CO), Crime (CR), Documentary (DO), Drama (DR), Fantasy (FA), IMAX (IM), Horror (HO), Mystery (MY), Musical (MU), Romance (RO), SciFi (SC), Thriller (TH), War (WA), Western (WE).}

\subsection{Comprehensive Evaluation}
\label{sec:control_evaluation}

In this section, we evaluate how modification of SAE activations influences genre presence in recommendations as well as recommendation metrics. We take top features, corresponding to Action, Children, Fantasy, Sci-Fi, and Western into our analysis.

\begin{figure*}[ht]
\centering

\begin{subfigure}{0.95\linewidth}
\begin{subfigure}{0.195\linewidth}
     \includegraphics[width=\textwidth]{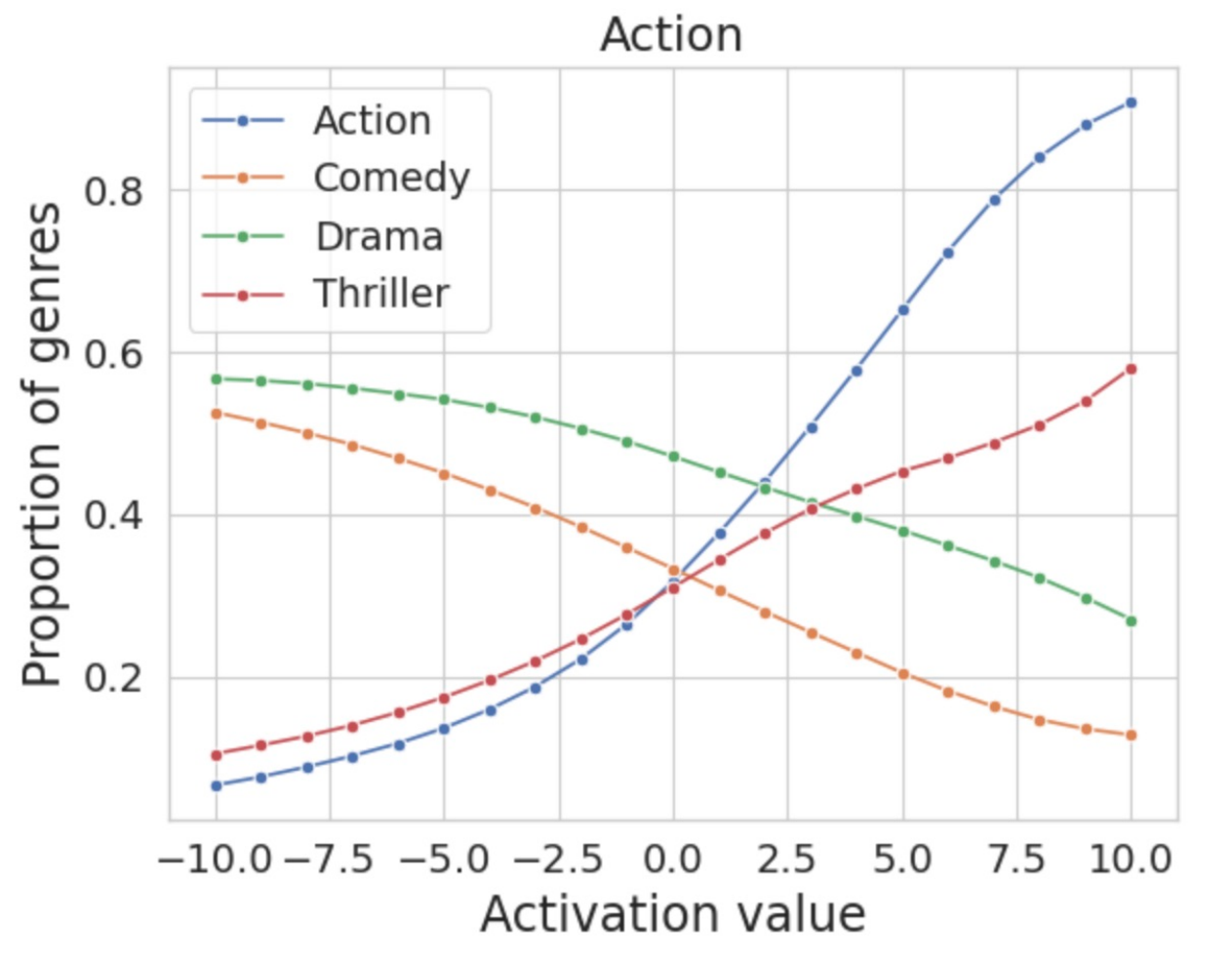}
 \end{subfigure}
 \begin{subfigure}{0.195\linewidth}
     \includegraphics[width=\textwidth]{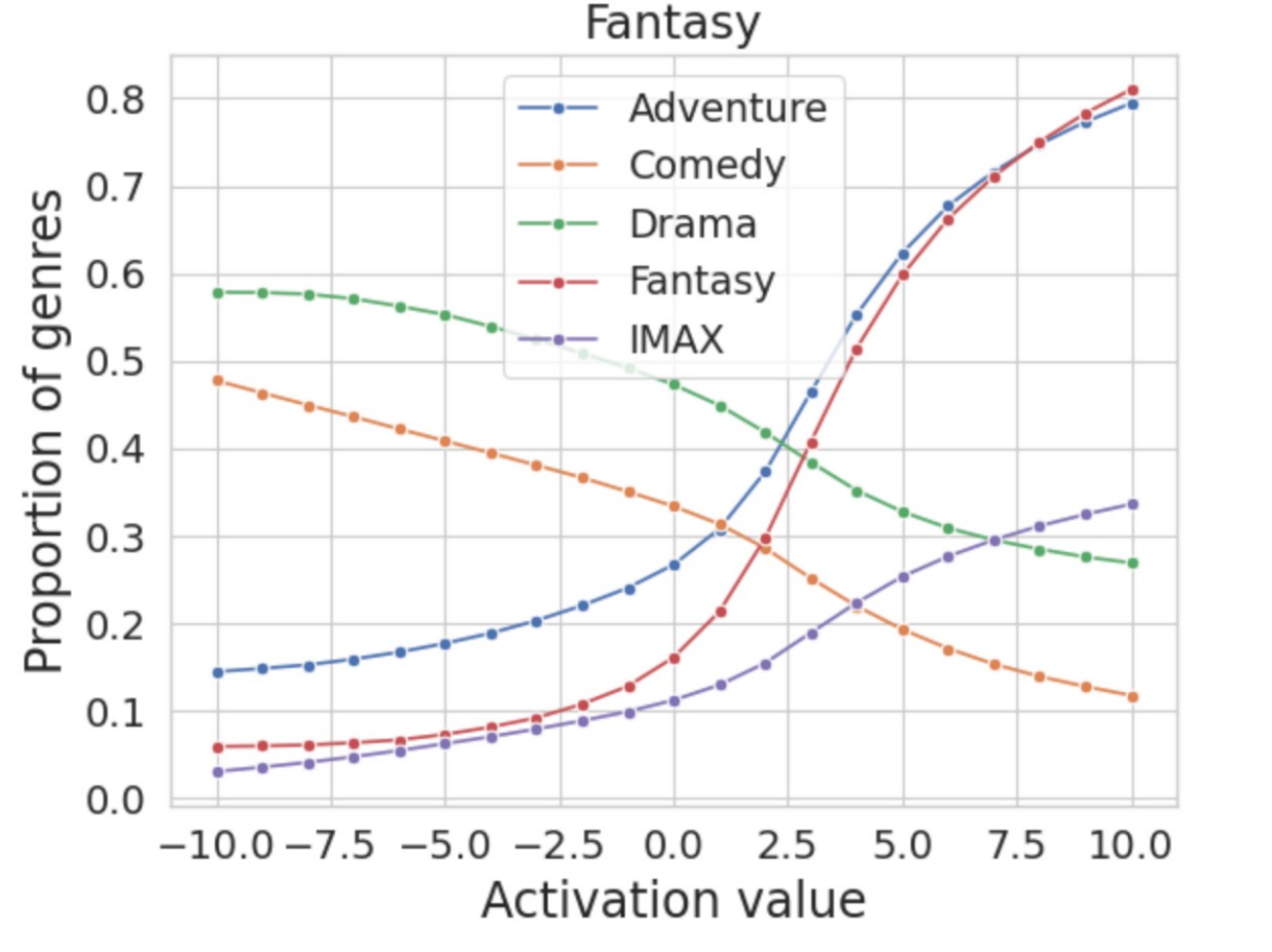}
 \end{subfigure}
 \begin{subfigure}{0.195\linewidth}
     \includegraphics[width=\textwidth]{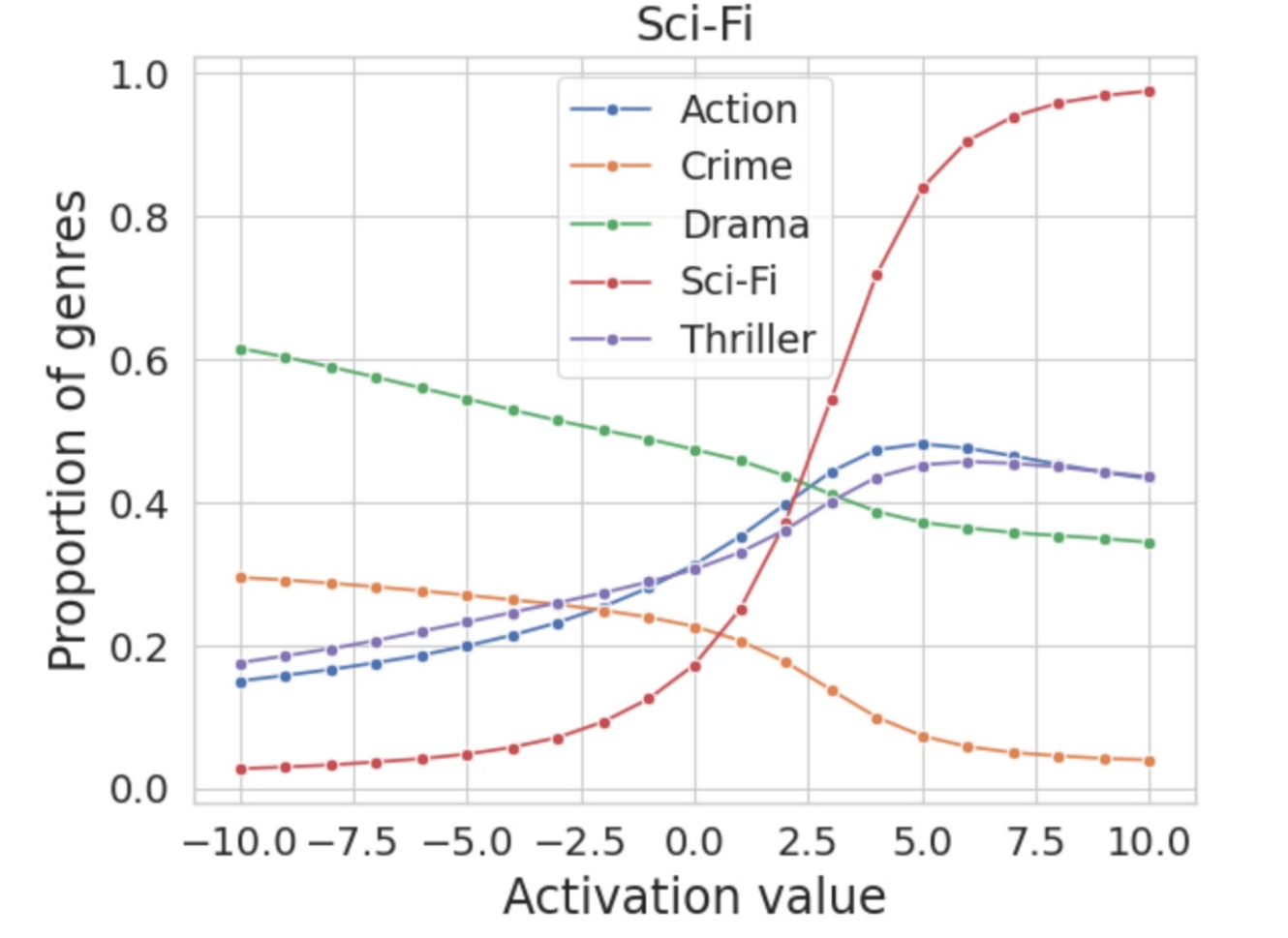}
 \end{subfigure}
 \begin{subfigure}{0.195\linewidth}
     \includegraphics[width=\textwidth]{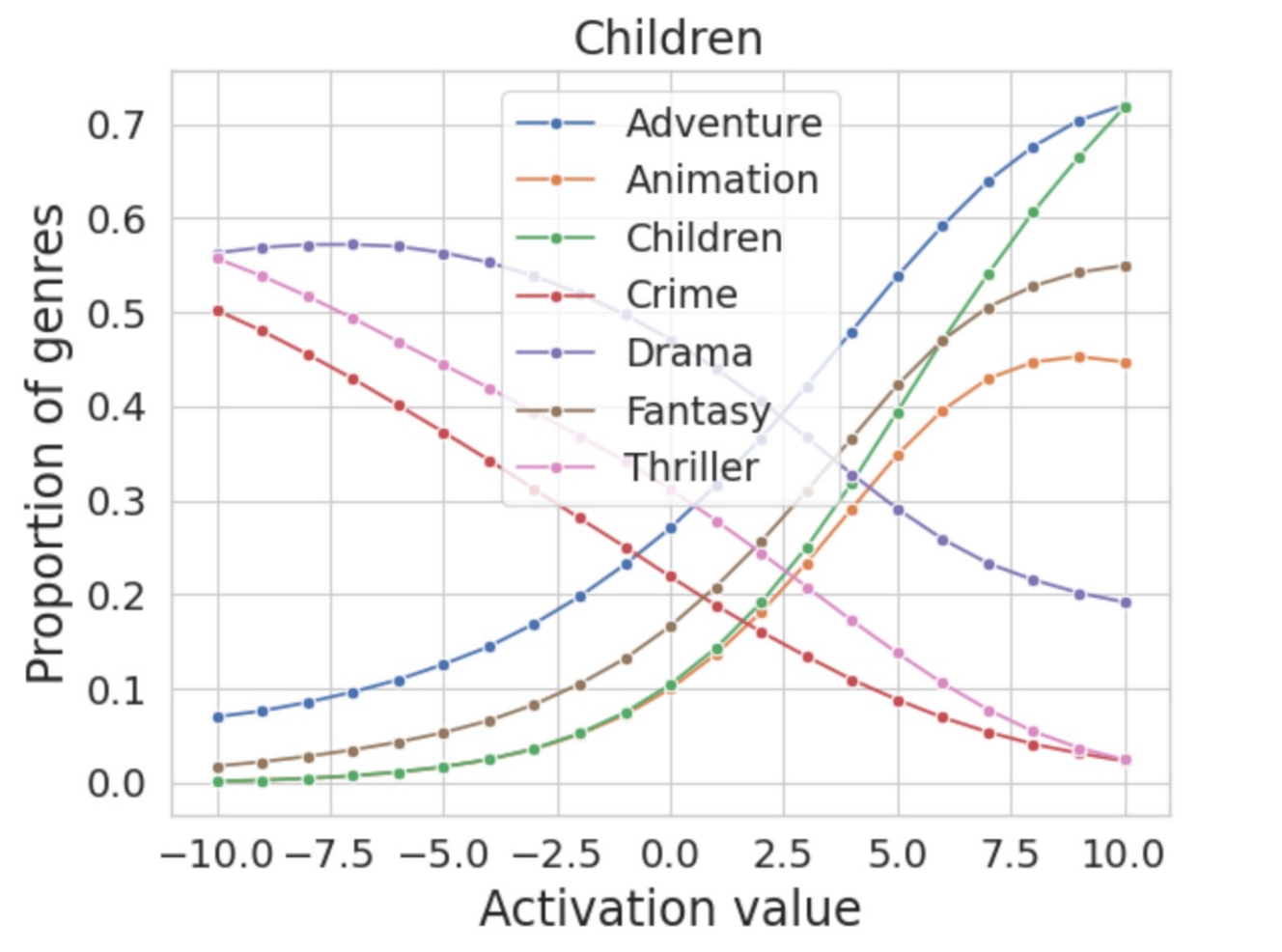}
 \end{subfigure}
 \begin{subfigure}{0.195\linewidth}
     \includegraphics[width=\textwidth]{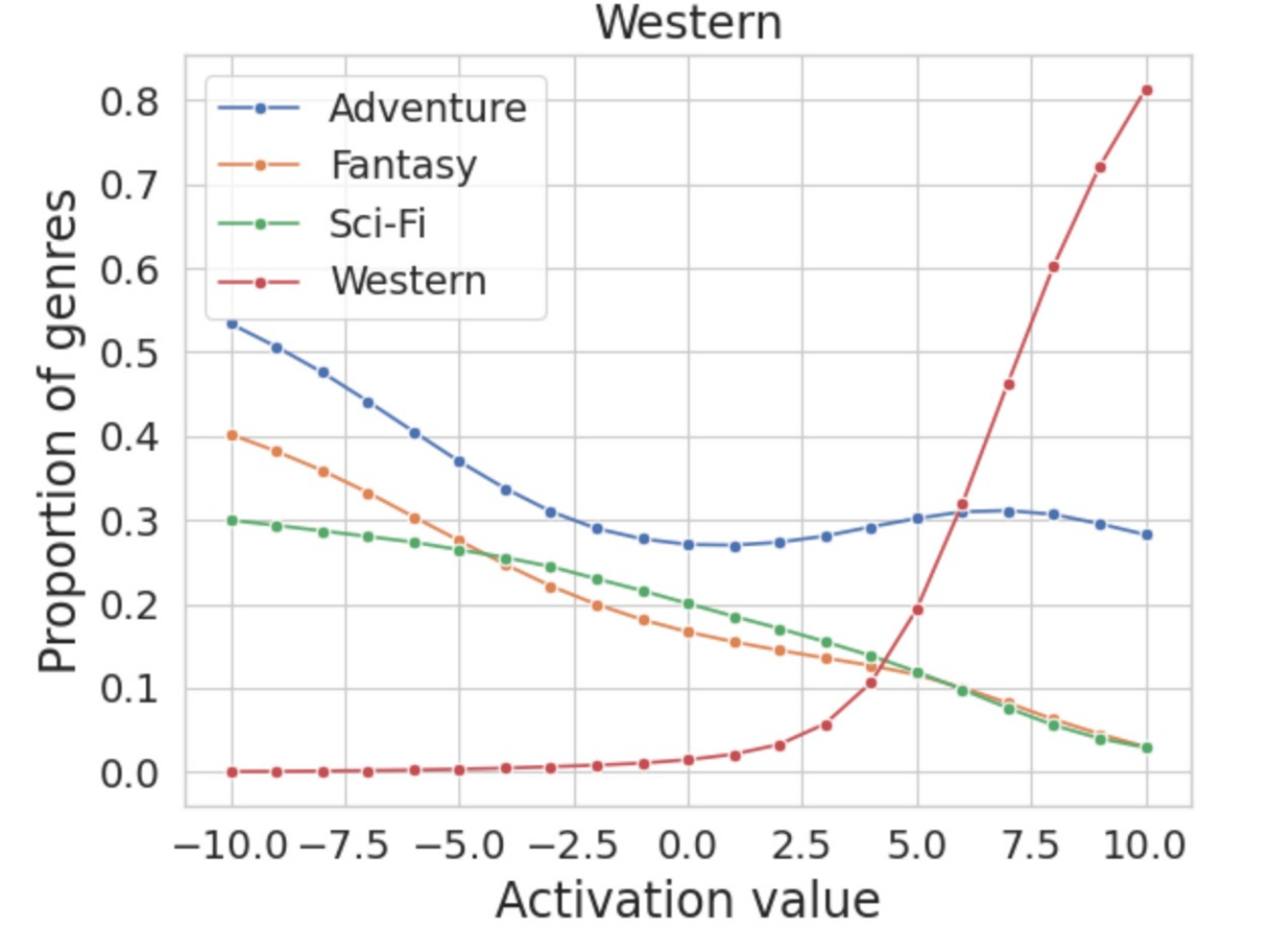}
 \end{subfigure}
 \caption{The proportion of genres in recommendations. Each plot corresponds to one single feature responsible for a given genre.}
 \end{subfigure}
 \resizebox{0.95\linewidth}{!}{
 \begin{subfigure}{0.33\textwidth}
     \includegraphics[width=\textwidth]{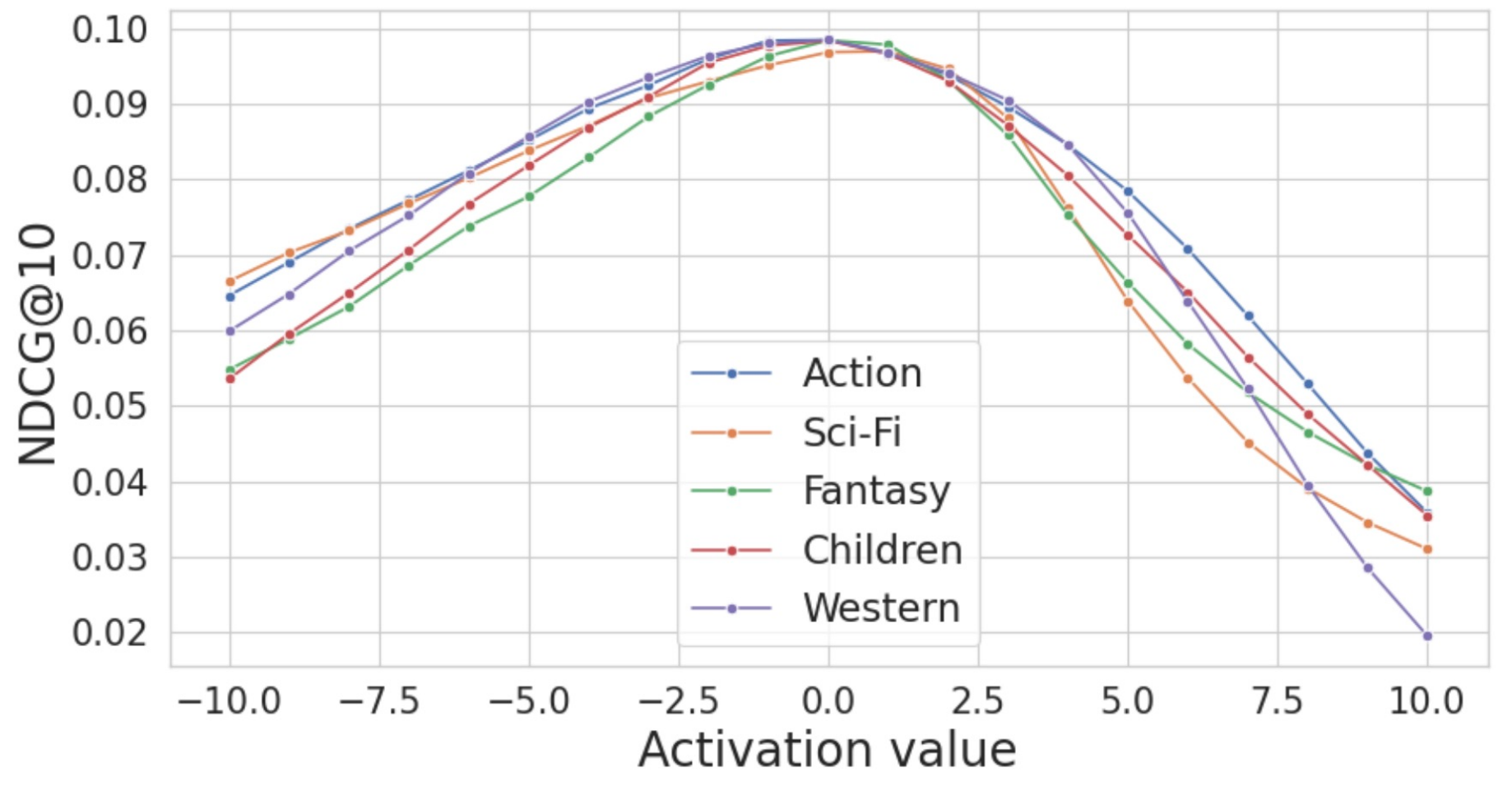}
     \caption{NDCG@10}
 \end{subfigure}
 \begin{subfigure}{0.33\textwidth}
     \includegraphics[width=\textwidth]{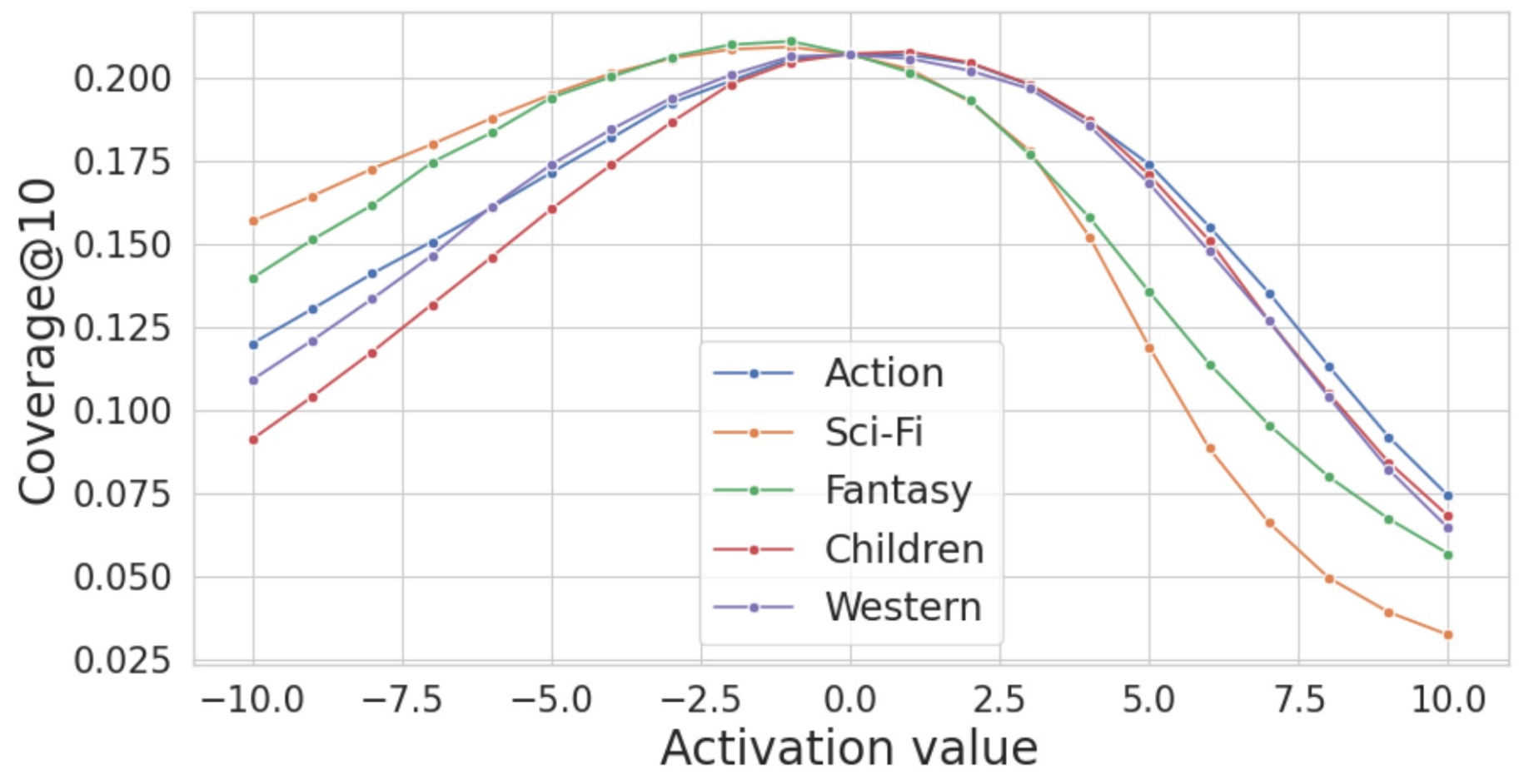}
     \caption{Coverage@10}
 \end{subfigure}
 \begin{subfigure}{0.33\textwidth}
     \includegraphics[width=\textwidth]{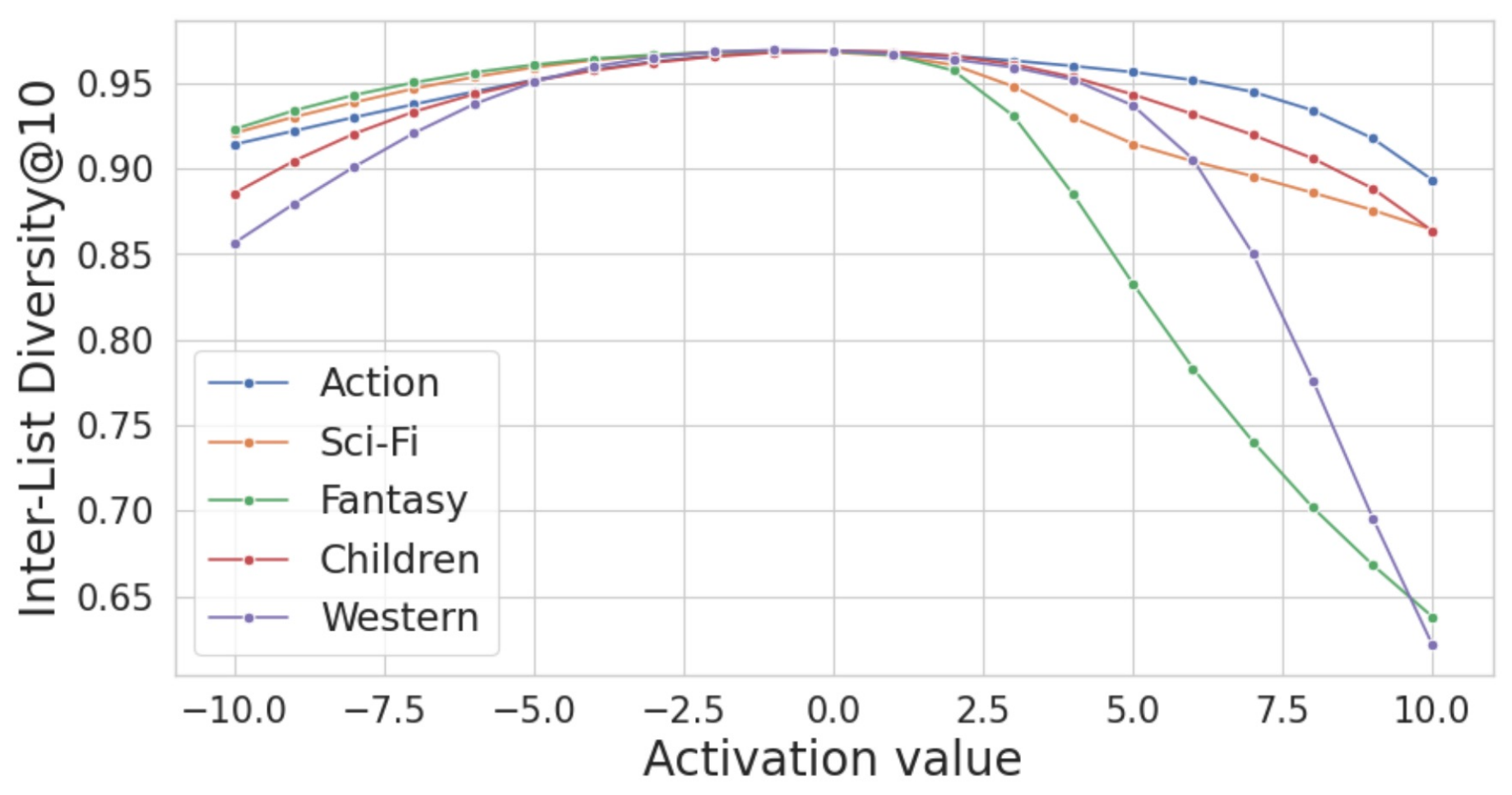}
     \caption{Inter-list Diversity@10}
 \end{subfigure}
 }
\caption{Effect of changing a SAE feature on the proportion of genres in recommendations (a) and the recommendation metrics(b-d). For (a), each curve represents genre proportions for different activation values. Only genres with a large proportion of change are shown for clarity. For (b-d), each curve corresponds to one single feature that is responsible for a given genre.}
\label{fig:intervention_genres}
\end{figure*}

\subsubsection{Proportion of genres}

For each feature, we perform an intervention, set its activation on the last sequence step to some value, and make a top 10 recommendation list for each user. Then, we compute the mean proportion of each genre in recommendations. This intervention is performed for activation values in the range [-10, 10] with step 1.

Figure~\ref{fig:intervention_genres} (a) shows the resulting dependency of genre proportions on the activation value of considered features. For all features, the proportion of the corresponding genre increases monotonically with activation value, is close to zero for large negative activations, and reaches very high values for large positive activations. It is clear that the steering of such features indeed allows us to control the presence of the desired genre in the recommendations.

The presence of other genres behaves in different ways. The proportion of strongly differing genres decreases (e.g. Drama, Thriller, and Crime for Children feature). The proportion of similar genres increases (e.g. Adventure for Fantasy feature). The Western feature increases only the Western genre. All these patterns correspond to patterns from Figure~\ref{fig:genre_corr_heatmap}, so the feature interpretations from Section~\ref{sec:interpretation} fit well with their downstream behavior.

\subsubsection{Quality of recommendations}

An important question is how this intervention affects the quality of recommendations. The model may recommend the selected genre, but the recommendations can become irrelevant to users. Or the model may lose personalization properties and recommend the same popular movies from the selected genre to all users. In this case, feature steering will become useless.

To check this, we compute the dependency of NDCG, Coverage, and Inter-list Diversity on feature activation values during the intervention (see Figure~\ref{fig:intervention_genres} (b-d)).
NDCG measures recommendation accuracy, while Coverage and Inter-list Diversity indicate the influence on diversity and personalization.
Inter-list Diversity is the average pairwise cosine distance between recommendation lists generated for different users, measuring how these lists are different from each other \cite{zhou2010solving,kaminskas2016diversity}.

As one might expected, the intervention led to some decrease in the recommendation metrics. The patterns are similar for different features. The intervention with activation absolute values less than 2 is safe, as the decrease in NDCG@10 and Coverage@10 is less than 10\% for most of the cases. Inter-list Diversity is also relatively stable in this range. Interventions with values greater than 5 should be performed with caution because metrics can decrease significantly. Interestingly, patterns for large negative values are similar to large positive values. When we suppress some genres, at the same time, we greatly increase the proportion of dissimilar genres, so the recommendations become less accurate and diverse.

The results of the intervention on the genre proportions and recommendation metrics for BERT4Rec and the Music4all dataset are provided in the appendix.

\subsubsection{Comparison with linear probing.}

\begin{figure}[ht!]
    \centering
    \includegraphics[width=0.9\linewidth, trim={0 0.2cm 0 1.6cm},clip ]{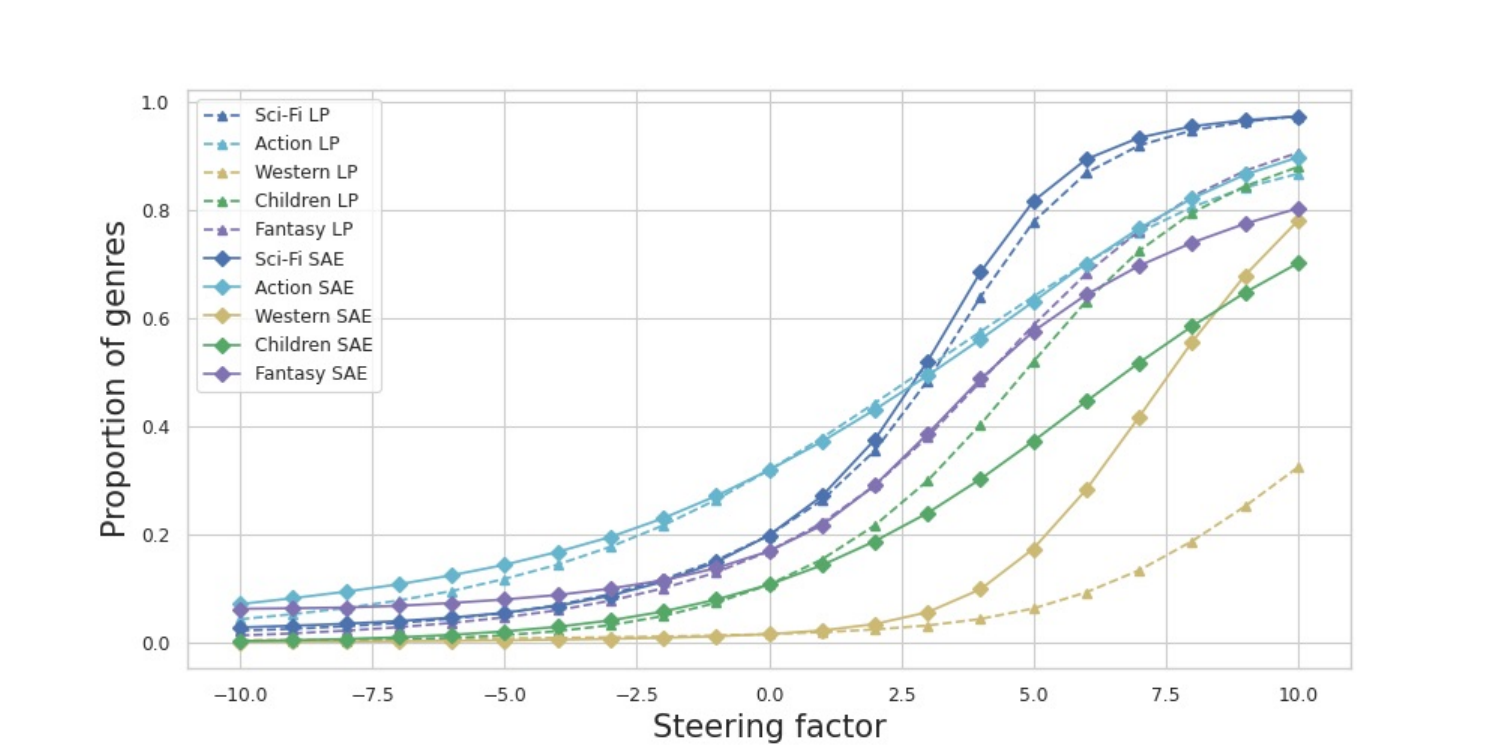}
    \caption{The proportion of the target genre in the recommendations. Comparison between control with SAE features and control with linear probing. The solid line corresponds to SAE, and the dashed line corresponds to linear probing.}
    \label{fig:linear_probing}
\end{figure}

Linear probing is a straightforward and widely used method to train a supervised classifier (the probe) on the activations of a network, used to find a direction that discriminates between some binary labels \cite{liInferenceTimeInterventionEliciting2023}. We follow this approach to evaluate how the performance of the unsupervised SAE compares to that of the supervised method. We compare SAE with linear probing (logistic regression, trained on the same inputs -- model activations -- as SAE, but with binary genre labels as targets) in the task of controlling the model behavior. Unsurprisingly, interpretability metrics are substantially higher for linear probing (mean correlation is 0.7), as it is a supervised method in contrast to fully unsupervised SAE approach. However, downstream effects turn out to be on par with SAE. Figure~\ref{fig:linear_probing} illustrates effect of adding a steering vector with different factors to activations of the original transformer layer. Both SAE and probing vectors are normalized to have unit variance. The influence of SAE is very similar to linear probing for Sci-Fi, Action and Fantasy, worse for Children and better for rare Western genre. The influence on recommendation metrics follow a similar pattern. The fact that the results of SAE are comparable with supervised linear probing proves that it is indeed a promising approach.

\section{Conclusion}

In this work, we extended sparse autoencoders to the domain of sequential recommendations. We demonstrated that SAEs can learn meaningful and interpretable features and introduced new interpretability metrics specifically tailored to sequential recommendation settings. Moreover, we showed that these learned features can be used to effectively and flexibly control model behavior, with their downstream effects aligning consistently with their interpretations.

Our findings highlight a new perspective on understanding and steering sequential recommender models. Beyond advancing interpretability, the proposed approach provides a practical mechanism for controllable personalization, enabling recommendation systems that are more transparent, adaptable, and user-aligned. Future work may explore alternative sparse autoencoder architectures, such as JumpReLU~\cite{rajamanoharan2024jumping} and TopK SAEs~\cite{gaoScalingEvaluatingSparse2024}, develop more refined control techniques, and investigate how interpretable latent representations can be integrated with broader recommendation objectives such as fairness, diversity, and long-term user satisfaction.

\bibliographystyle{ACM-Reference-Format}
\bibliography{content/bibliography}

\clearpage
\FloatBarrier
\appendix

\section{Additional Results for BERT4Rec}
\label{sec:appendix_bert4rec}

\begin{figure}[H]
    \centering
    \includegraphics[width=0.9\linewidth]{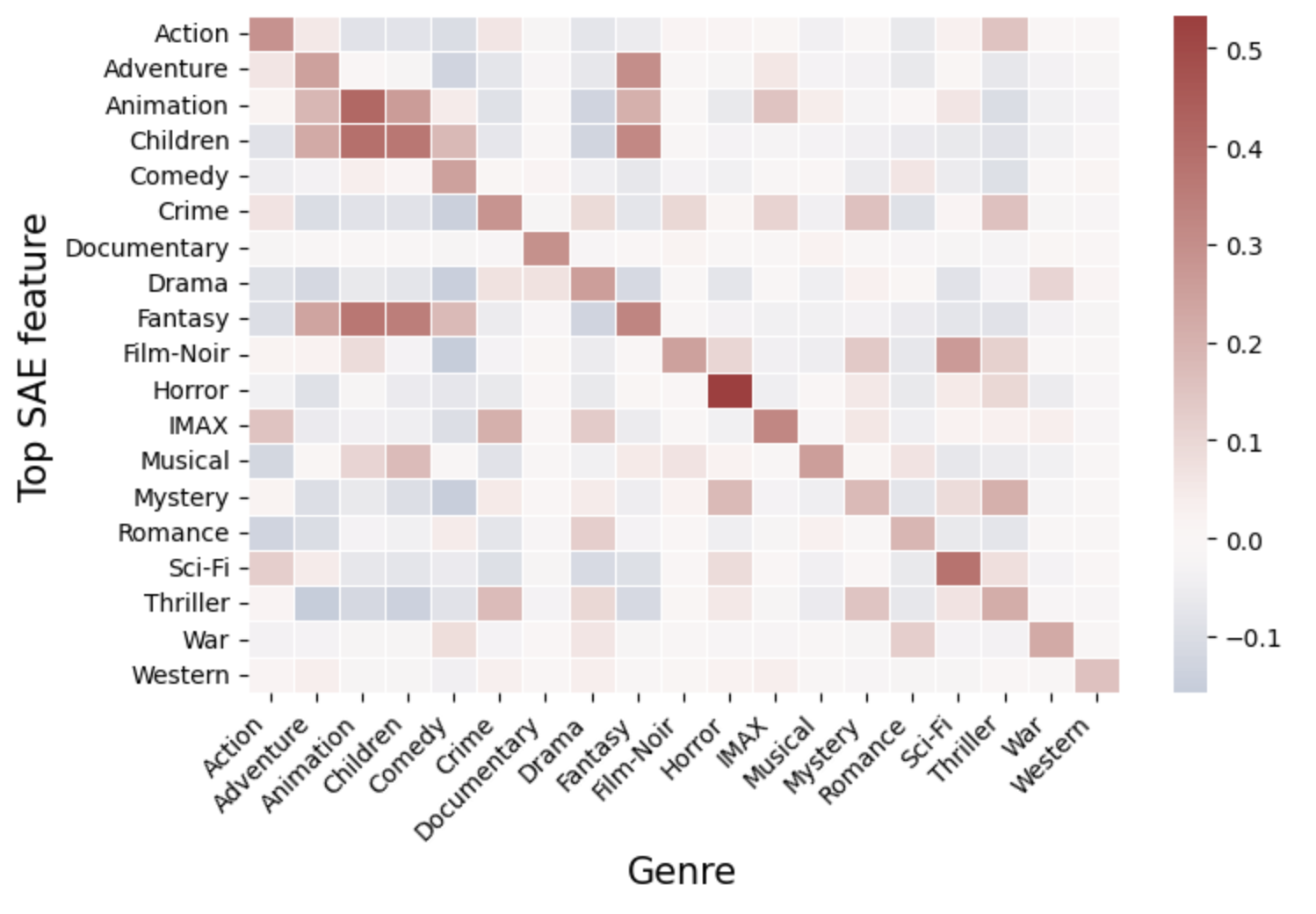}
    \caption{Correlation between genres and top feature (the feature with maximum correlation) for each genre for BERT4Rec on the Movielens-20M dataset. One row corresponds to one feature and contains its correlations with all genres.}
    \label{fig:genre_corr_heatmap_bert4rec}
\end{figure}

For BERT4Rec, we perform the full set of experiments on the Movielens-20M dataset. First, we compute correlation, ROC AUC, and sensitivity between each SAE feature and genre and identify the top feature with the highest correlation, following the procedure from Section~\ref{sec:finding_top_neurons}.
Table~\ref{tab:metrics_by_genre_bert_ml20m} presents the metric values for these top features. Similar to GPTRec, the genres Horror, Animation, Sci-Fi, and Children are among those with the highest correlations. Overall, the absolute metric values are lower than for GPTRec.
Figure~\ref{fig:genre_corr_heatmap_bert4rec} shows the correlations of these top features with all genres. As before, the features tend to be monosemantic, and the observed patterns align well with genre semantics.

\begin{figure}[H]
    \centering
    \includegraphics[width=\linewidth]{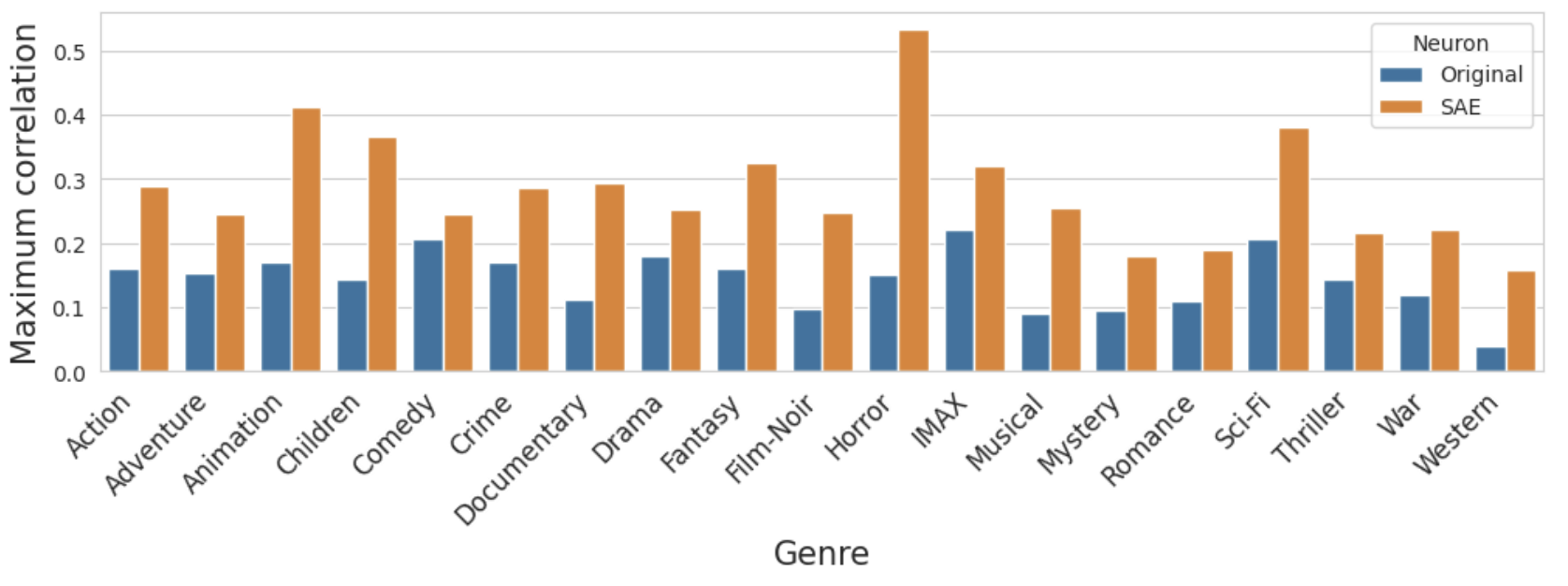}
    \caption{Comparison between the interpretability of SAE features and neurons of the original transformer layer for BERT4Rec on the Movielens-20M dataset. The maximum correlation for each genre is blue for the transformer layer and orange for SAE.}
    \label{fig:corr_original_vs_sae_bert4rec}
\end{figure}

Next, we perform the analysis from Section~\ref{sec:original_vs_sae}, comparing the interpretability of SAE features with that of the original transformer layer. Figure~\ref{fig:corr_original_vs_sae_bert4rec} displays the correlation values of the top transformer neurons and SAE features for each genre. Although the absolute correlations are lower than those for GPTRec, both for SAE and transformer neurons, the SAE features remain consistently more interpretable than the original transformer neurons.

\nopagebreak[4]

\begin{table}[H]
\caption{Metrics for SAE features with maximum correlation for each genre for BERT4Rec on the Moivelens-20M dataset. The last row shows the averages across all genres, which are considered as overall interpretability metrics.}
\label{tab:metrics_by_genre_bert_ml20m}
\centering
\small
\resizebox{\columnwidth}{!}{
\begin{tabular}{@{}lrrrr@{}}
\toprule
\textbf{Genre}       & \textbf{Correlation} & \textbf{ROC AUC} & \textbf{Sensitivity} & \textbf{Genre popularity}  \\ 
\midrule
Horror & 0.533 & 0.798 & 0.681 & 0.066 \\ 
Animation & 0.412 & 0.702 & 0.572 & 0.083 \\ 
Sci-Fi & 0.380 & 0.687 & 0.577 & 0.182 \\ 
Children & 0.367 & 0.630 & 0.543 & 0.095 \\ 
Fantasy & 0.326 & 0.621 & 0.497 & 0.135 \\ 
IMAX & 0.320 & 0.760 & 0.767 & 0.061 \\ 
Documentary & 0.293 & 0.751 & 0.721 & 0.011 \\ 
Action & 0.288 & 0.626 & 0.536 & 0.331 \\ 
Crime & 0.285 & 0.657 & 0.508 & 0.195 \\ 
Musical & 0.254 & 0.735 & 0.597 & 0.038 \\ 
Drama & 0.253 & 0.628 & 0.455 & 0.470 \\ 
Film-Noir & 0.247 & 0.775 & 0.747 & 0.011 \\ 
Adventure & 0.246 & 0.618 & 0.473 & 0.255 \\ 
Comedy & 0.246 & 0.600 & 0.436 & 0.311 \\ 
War & 0.221 & 0.628 & 0.488 & 0.061 \\ 
Thriller & 0.217 & 0.611 & 0.459 & 0.278 \\ 
Romance & 0.190 & 0.602 & 0.482 & 0.175 \\ 
Mystery & 0.179 & 0.625 & 0.535 & 0.099 \\ 
Western & 0.158 & 0.642 & 0.520 & 0.015 \\ 
\midrule
Mean & 0.285 & 0.668 & 0.558 & 0.151 \\ 
\bottomrule
\end{tabular}
}
\end{table}
\nopagebreak[4]

\begin{figure*}[ht]
\centering
\begin{subfigure}{\linewidth}
\begin{subfigure}{0.195\linewidth}
     \includegraphics[width=\textwidth]{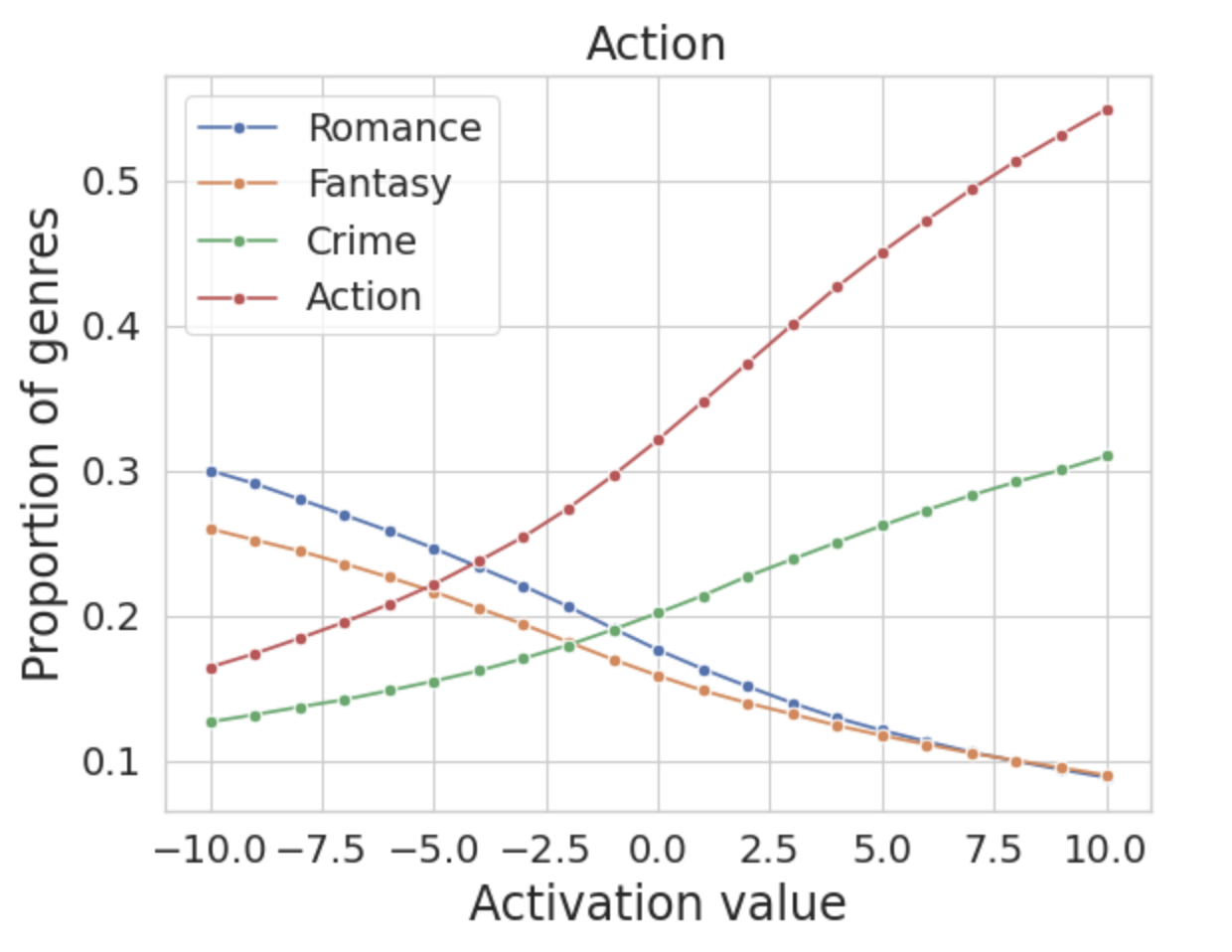}
 \end{subfigure}
 \begin{subfigure}{0.195\linewidth}
     \includegraphics[width=\textwidth]{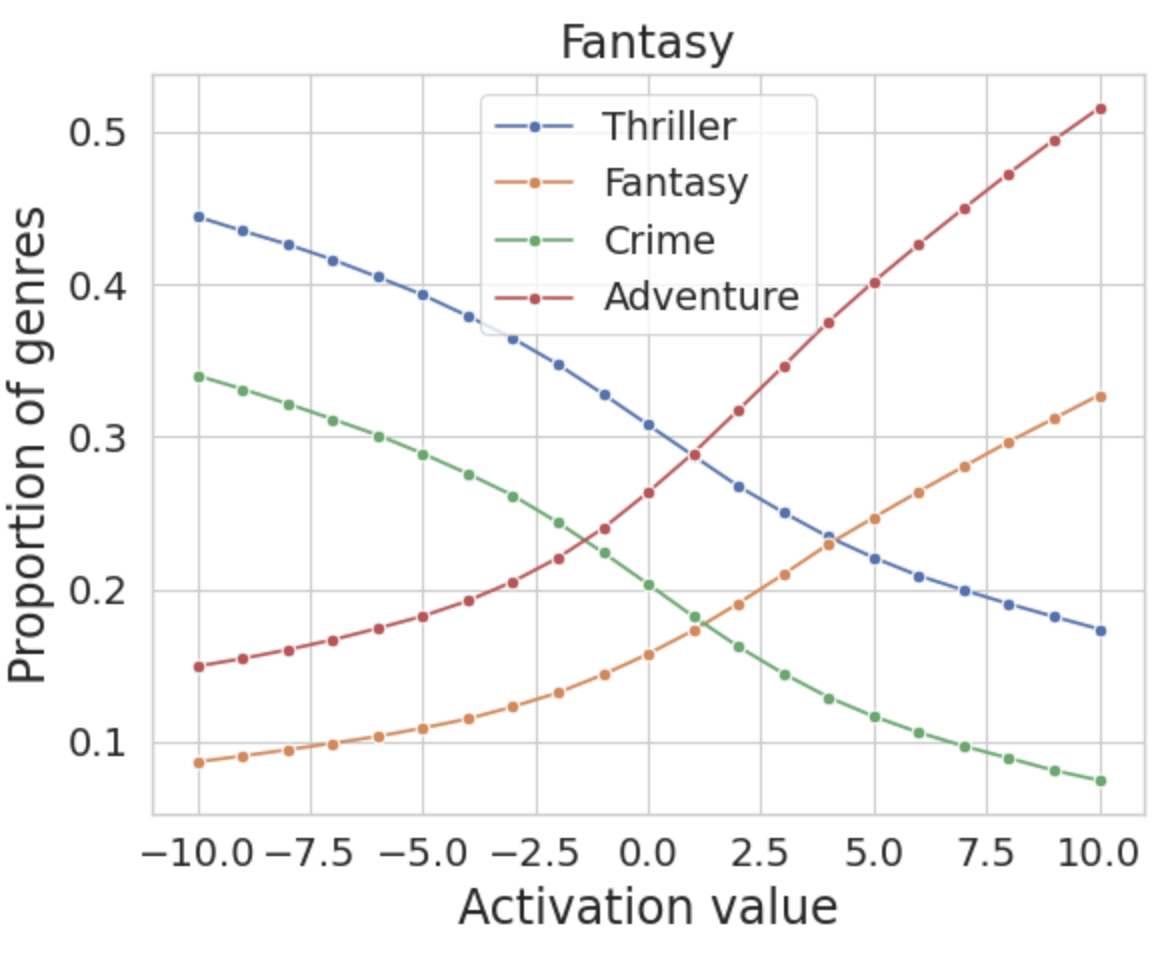}
 \end{subfigure}
 \begin{subfigure}{0.195\linewidth}
     \includegraphics[width=\textwidth]{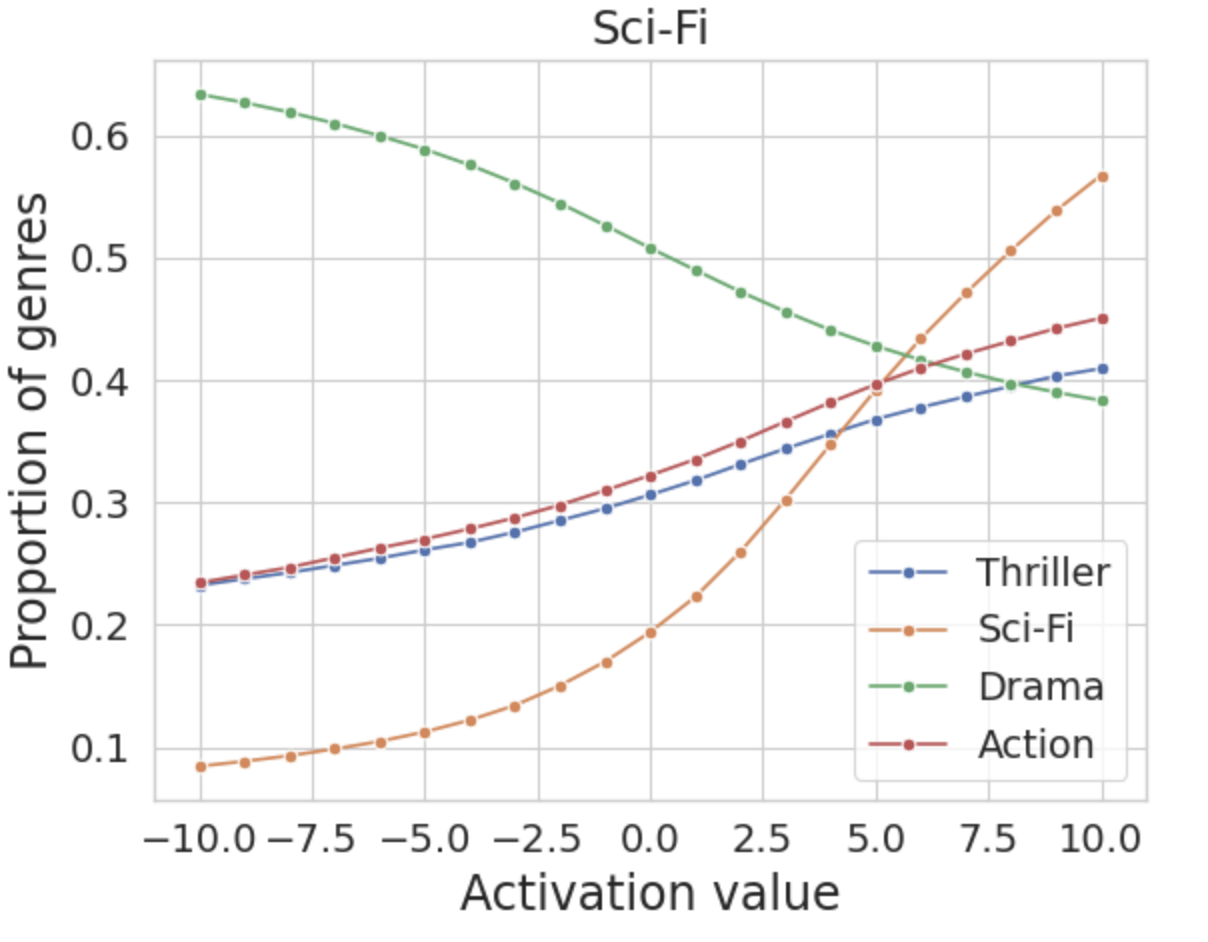}
 \end{subfigure}
 \begin{subfigure}{0.195\linewidth}
     \includegraphics[width=\textwidth]{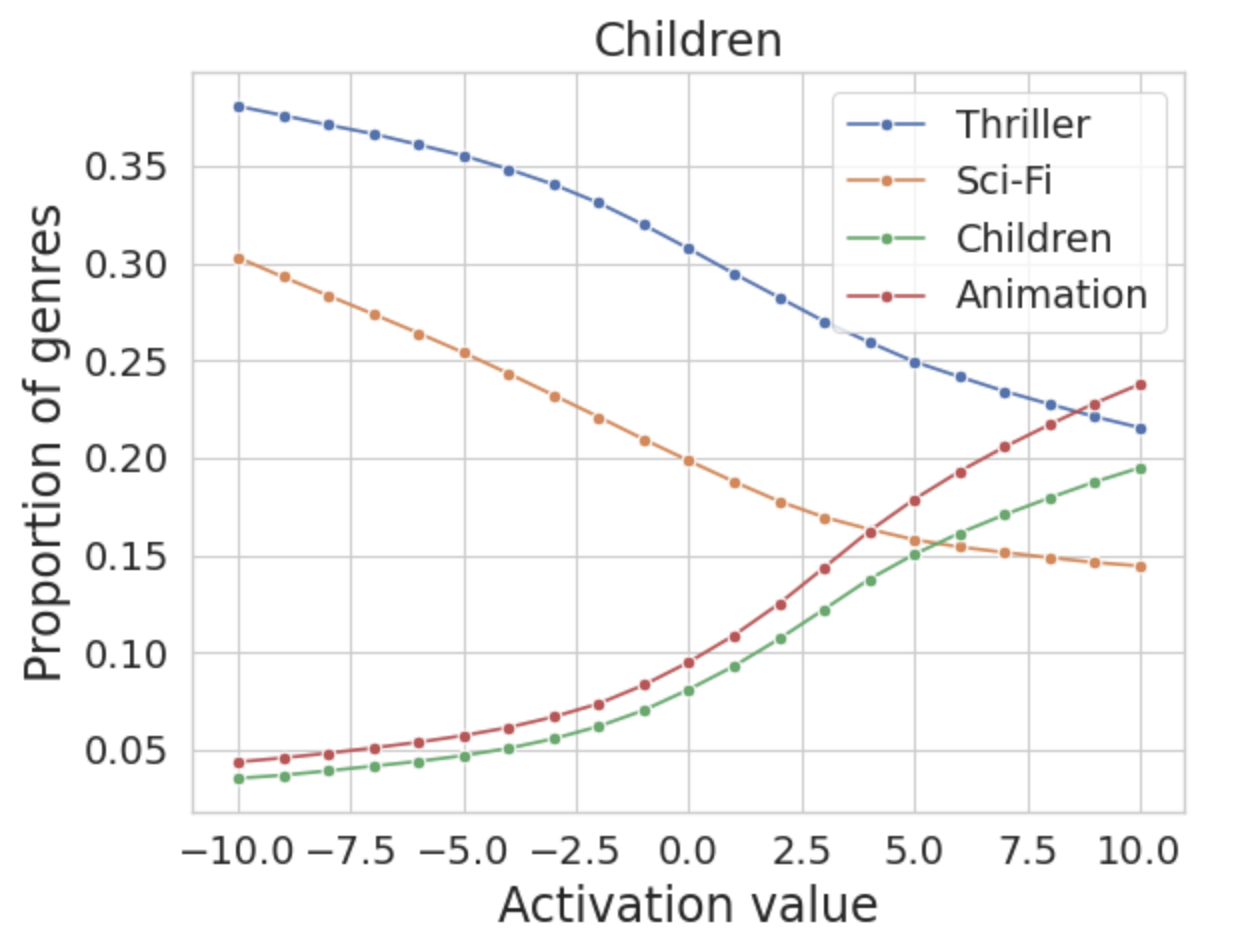}
 \end{subfigure}
 \begin{subfigure}{0.195\linewidth}
     \includegraphics[width=\textwidth]{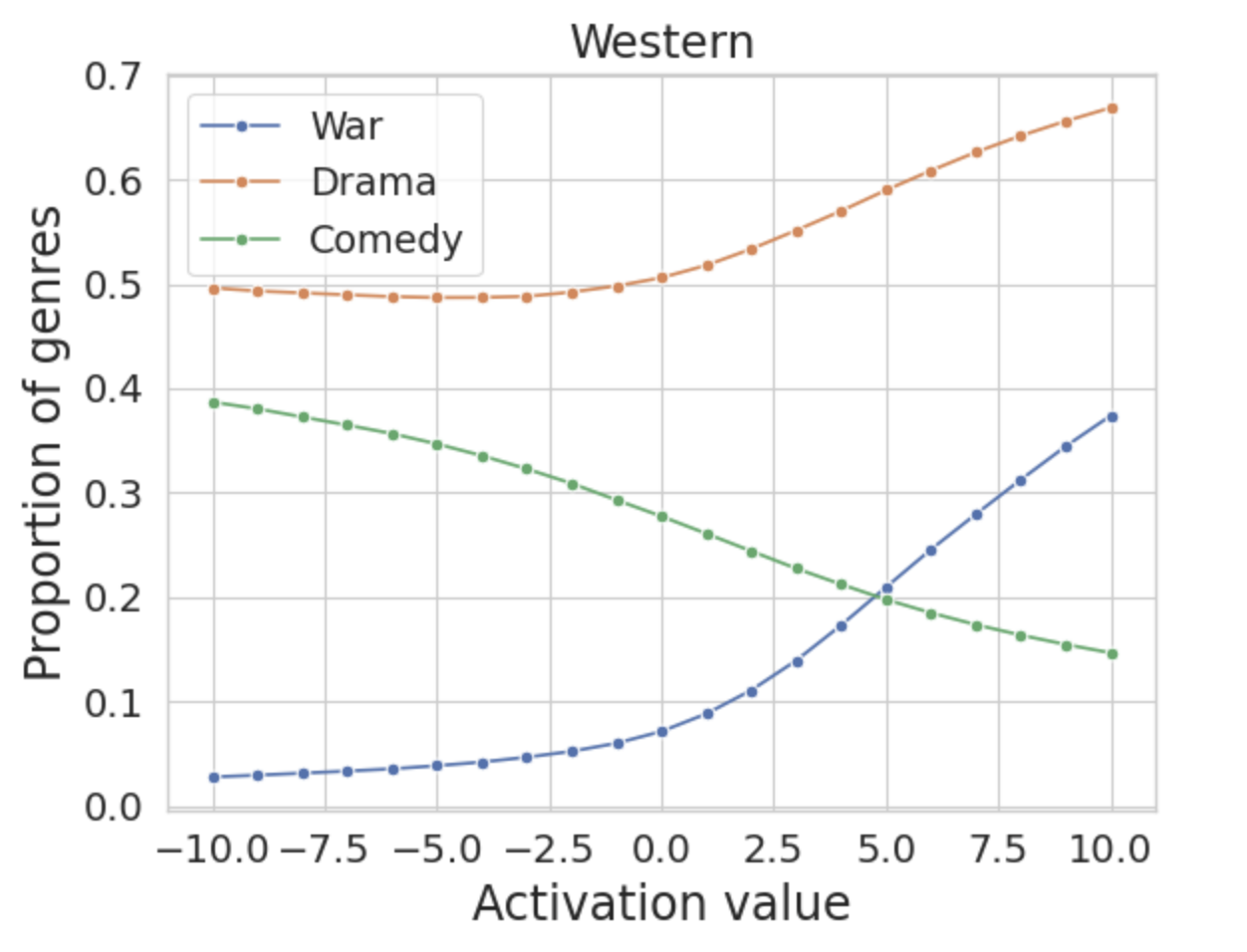}
 \end{subfigure}
 \caption{The proportion of genres in recommendations. Each plot corresponds to one single feature responsible for a given genre.}
 \end{subfigure}
 \begin{subfigure}{0.33\textwidth}
     \includegraphics[width=\textwidth]{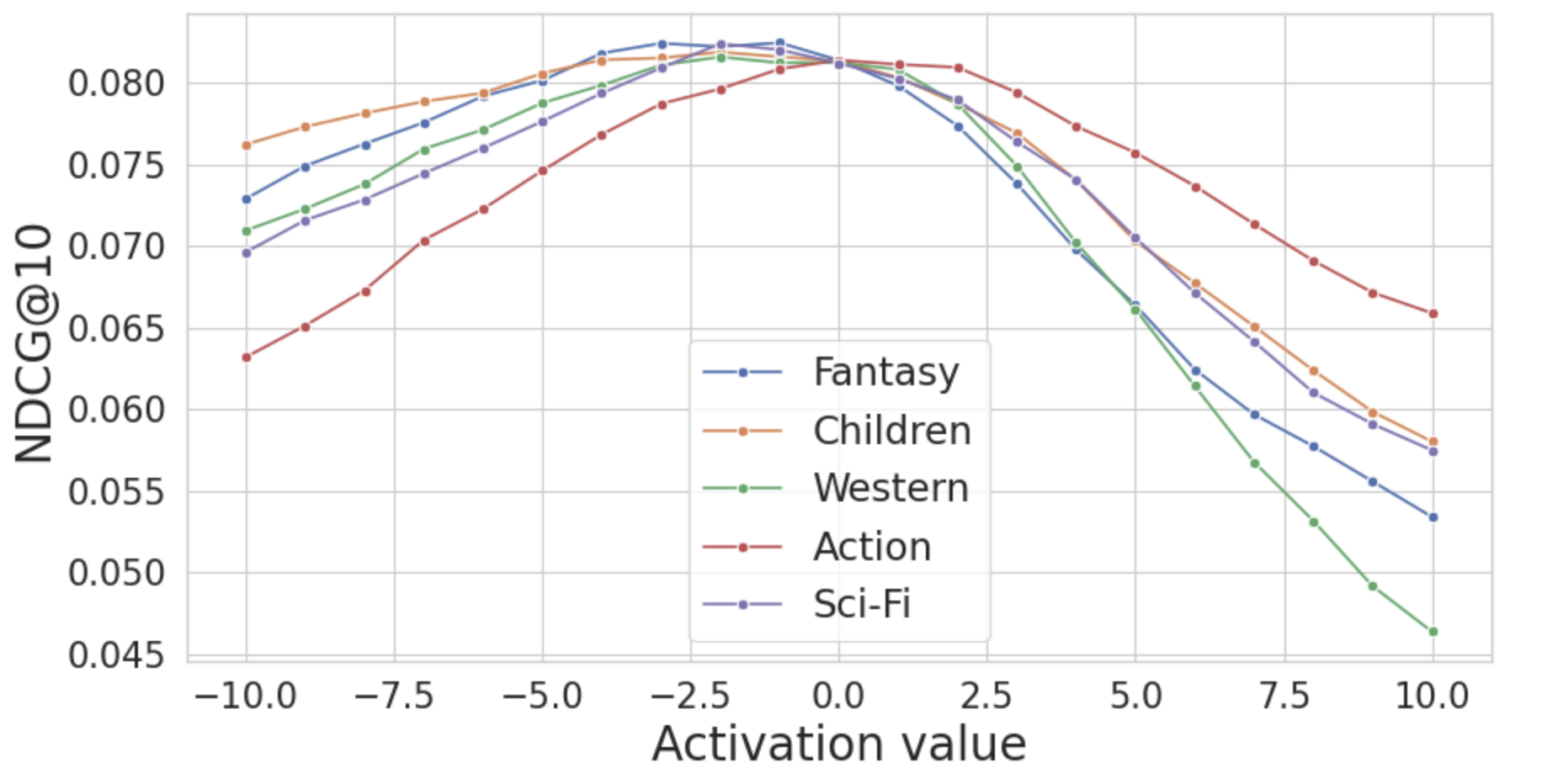}
     \caption{NDCG@10}
 \end{subfigure}
 \begin{subfigure}{0.33\textwidth}
     \includegraphics[width=\textwidth]{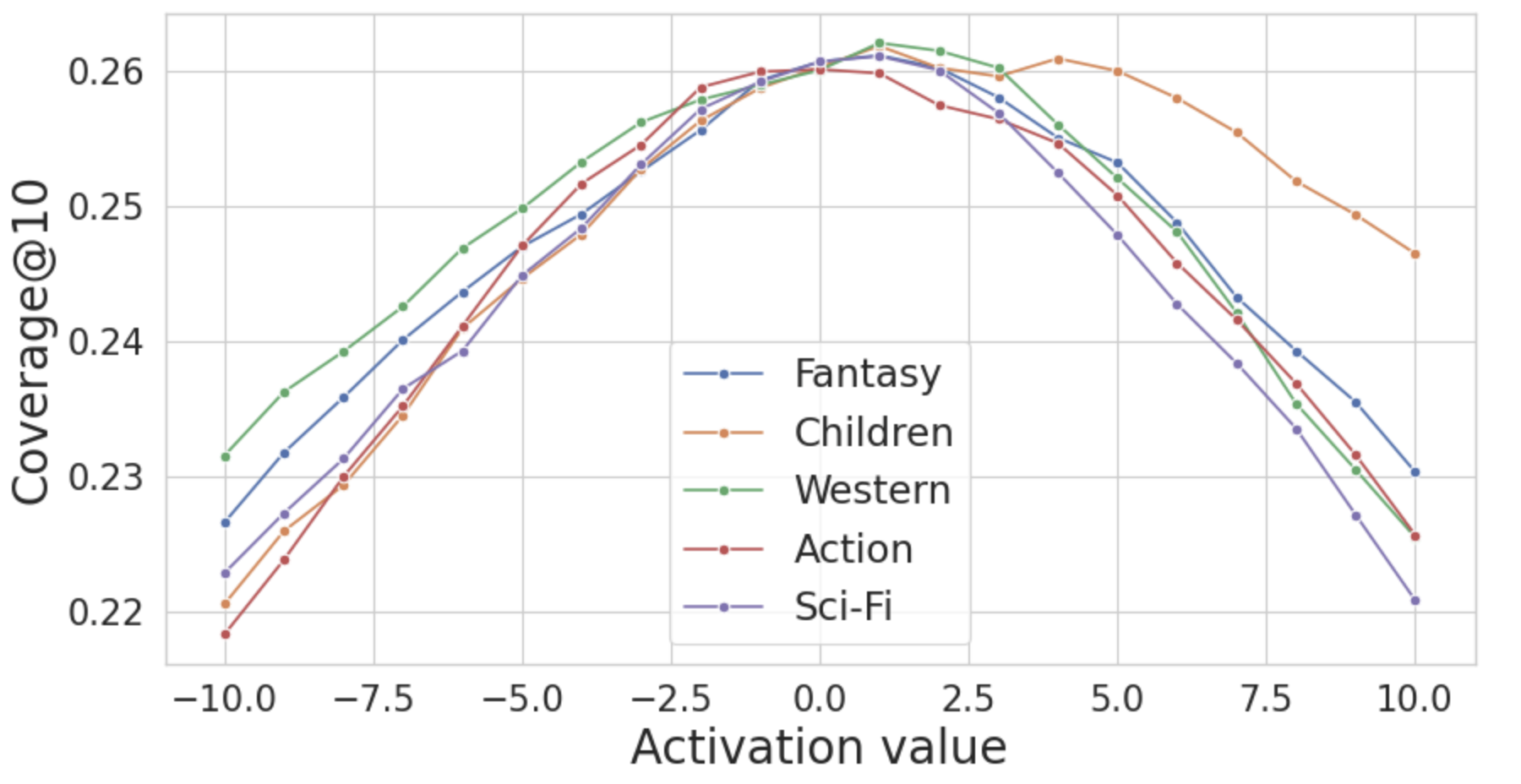}
     \caption{Coverage@10}
 \end{subfigure}
 \begin{subfigure}{0.33\textwidth}
     \includegraphics[width=\textwidth]{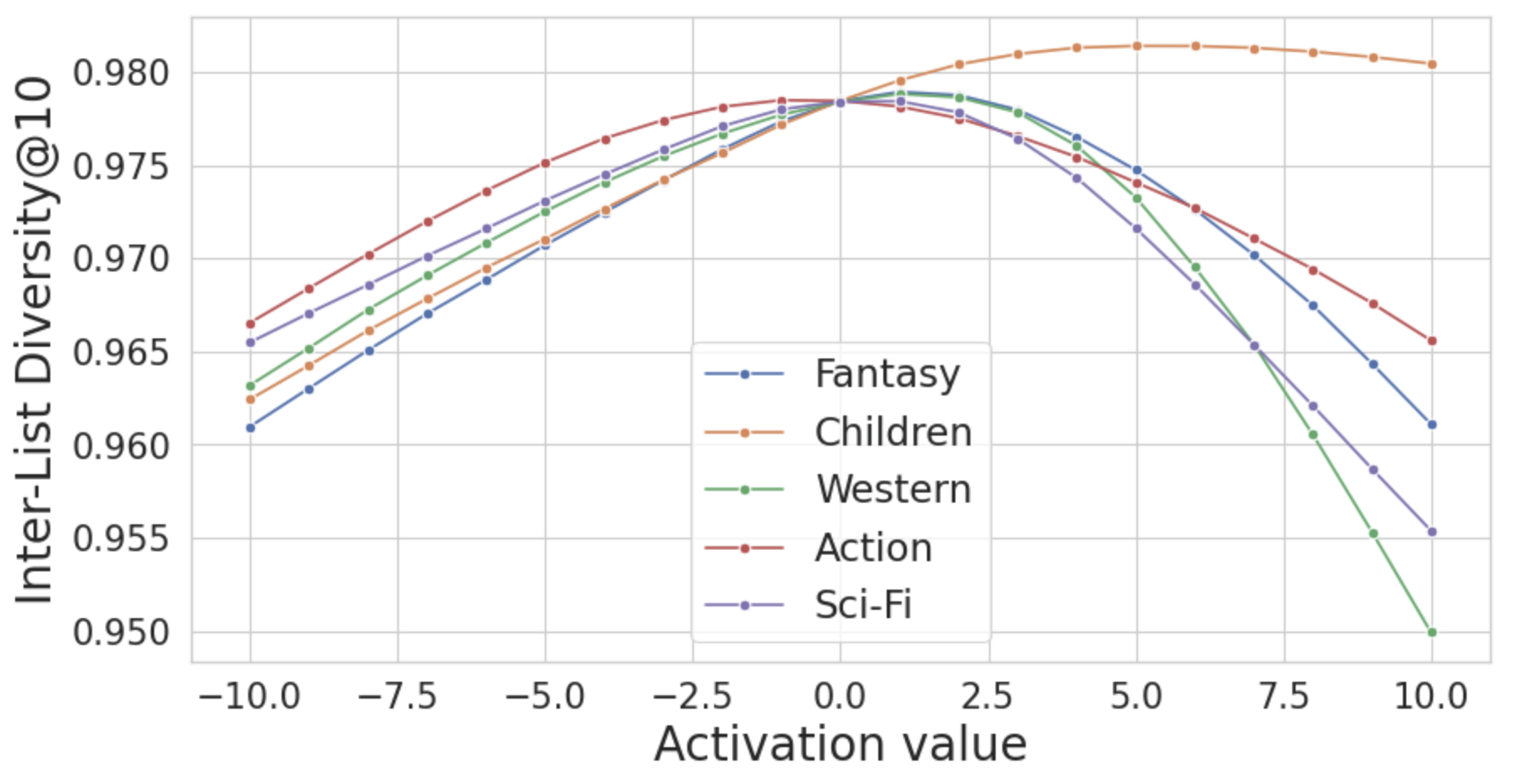}
     \caption{Inter-list Diversity@10}
 \end{subfigure}
\caption{Effect of changing a SAE feature on the proportion of genres in recommendations (a) and the recommendation metrics(b-d) for BERT4Rec on the Movielens-20M dataset. For (a), each curve represents genre proportions for different activation values. Only genres with a large proportion of change are shown for clarity. For (b-d), each curve corresponds to one single feature that is responsible for a given genre.}
\label{fig:intervention_genres_bert4rec}
\end{figure*}

Finally, we conduct extensive experiments on model control, as described in Section~\ref{sec:control_evaluation}. Figure~\ref{fig:intervention_genres_bert4rec} illustrates the effect of modifying a single SAE feature on the genre proportions and recommendation metrics. For Action, Sci-Fi, Fantasy, and Children, the genre proportions change as expected. Modifying the Western feature leads to a significant increase in the War genre, while Western itself is not strongly affected. This suggests that the feature is in fact associated with War, and that no meaningful feature was learned for Western. This is supported by the very low metric values for Western shown in Table~\ref{tab:metrics_by_genre_bert_ml20m}.

Regarding the effect on recommendation quality, the trend is similar to GPTRec: interventions with activation values below 2 result in small changes to the metrics, while stronger interventions can lead to more noticeable degradation in performance.

\nopagebreak[4]

\section{Additional Results for the Music4all Dataset}
\label{sec:appendix_music4all}

For the Music4All dataset, we perform the full set of experiments using GPTRec. Following the procedure from Section~\ref{sec:finding_top_neurons}, we compute correlation, ROC AUC, and sensitivity between each SAE feature and genre and identify the top feature with the highest correlation. Table~\ref{tab:metrics_by_genre_gpt_music4all} presents the metric values for these top features. The feature corresponding to k-pop shows particularly high values across all metrics, indicating strong and specific alignment with this genre. While other features show lower values, many still reach levels comparable to those observed on the Movielens-20M dataset.

\begin{figure}[htb]
    \centering
    \includegraphics[width=\linewidth]{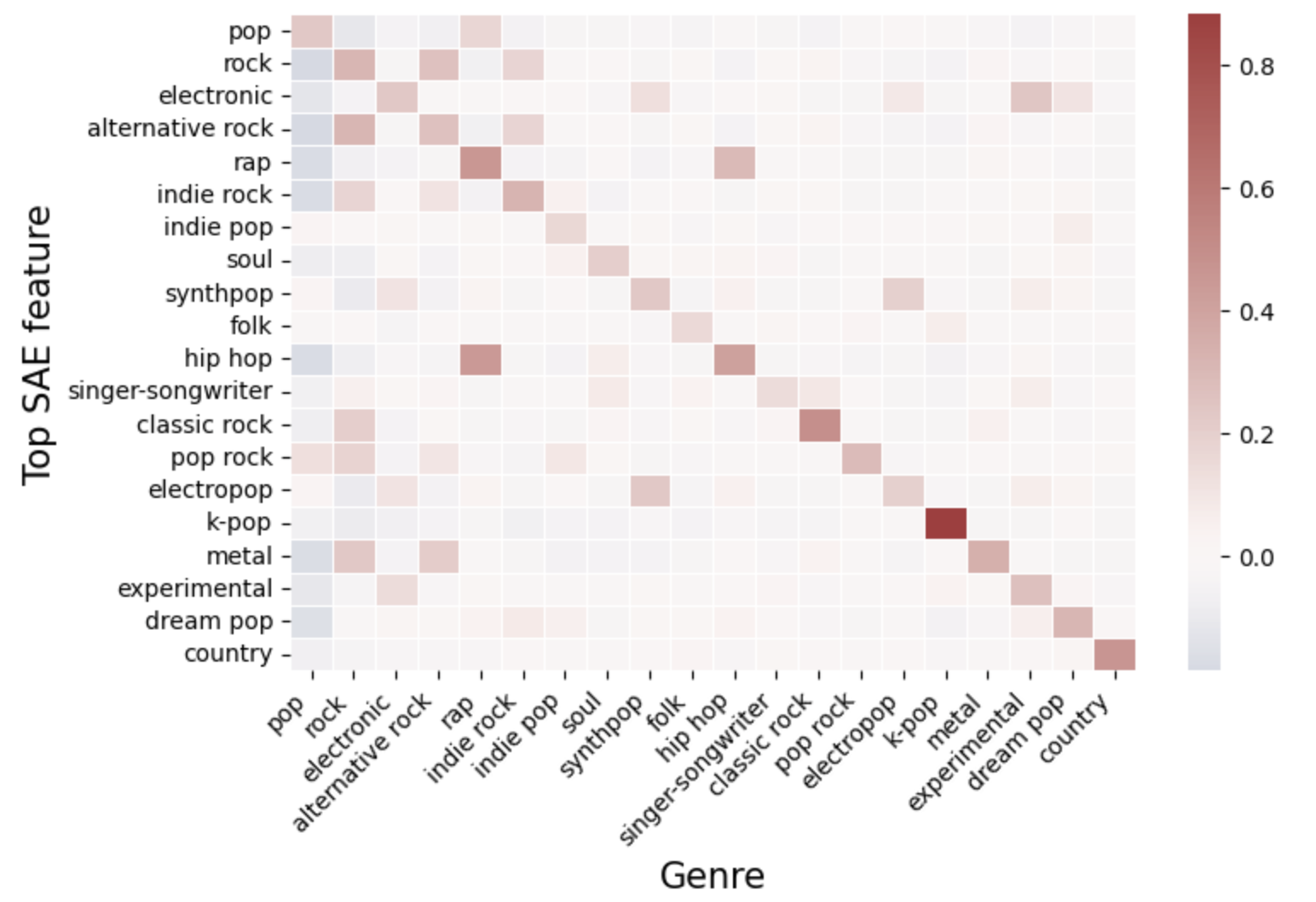}
    \caption{Correlation between genres and top feature (the feature with maximum correlation) for each genre for GPTRec on the Music4all dataset. One row corresponds to one feature and contains its correlations with all genres.}
    \label{fig:genre_corr_heatmap_music4all}
\end{figure}

Figure~\ref{fig:genre_corr_heatmap_music4all} shows the correlations of these top features with all genres. The features appear even more monosemantic than in the movie domain. Several semantically meaningful patterns emerge: the rock feature correlates with alternative rock and indie rock, the rap feature with hip-hop, the synthpop feature with electropop, and the metal feature with both rock and alternative rock.

\begin{table}[H]
\caption{Metrics for SAE features with maximum correlation for each genre for GPTRec on the Music4all dataset. The last row shows the averages across all genres, which are considered as overall interpretability metrics.}
\label{tab:metrics_by_genre_gpt_music4all}
\centering
\small
\resizebox{\columnwidth}{!}{
\begin{tabular}{@{}lrrrr@{}}
\toprule
\textbf{Genre}       & \textbf{Correlation} & \textbf{ROC AUC} & \textbf{Sensitivity} & \textbf{Genre popularity} \\ 
\midrule
k-pop & 0.883 & 0.983 & 0.971 & 0.033 \\
classic rock & 0.490 & 0.787 & 0.613 & 0.027 \\
country & 0.466 & 0.711 & 0.559 & 0.024 \\
rap & 0.457 & 0.723 & 0.546 & 0.070 \\
hip hop & 0.411 & 0.755 & 0.574 & 0.040 \\
metal & 0.342 & 0.815 & 0.719 & 0.022 \\
indie rock & 0.320 & 0.695 & 0.488 & 0.068 \\
dream pop & 0.317 & 0.735 & 0.531 & 0.022 \\
rock & 0.314 & 0.640 & 0.411 & 0.191 \\
pop rock & 0.288 & 0.619 & 0.381 & 0.033 \\
experimental & 0.268 & 0.656 & 0.398 & 0.025 \\
alternative rock & 0.258 & 0.673 & 0.422 & 0.069 \\
pop & 0.233 & 0.616 & 0.343 & 0.448 \\
synthpop & 0.230 & 0.635 & 0.402 & 0.046 \\
electronic & 0.227 & 0.596 & 0.314 & 0.107 \\
soul & 0.206 & 0.602 & 0.369 & 0.055 \\
electropop & 0.201 & 0.632 & 0.375 & 0.033 \\
indie pop & 0.162 & 0.584 & 0.308 & 0.057 \\
folk & 0.159 & 0.628 & 0.439 & 0.039 \\
singer-songwriter & 0.139 & 0.624 & 0.421 & 0.036 \\
\midrule
Mean & 0.331 & 0.685 & 0.479 & 0.072 \\
\bottomrule
\end{tabular}
}
\end{table}
\nopagebreak[4]

\nopagebreak[4]

\begin{figure*}[!htb]
\centering
\begin{subfigure}{\linewidth}
\begin{subfigure}{0.195\linewidth}
     \includegraphics[width=\textwidth]{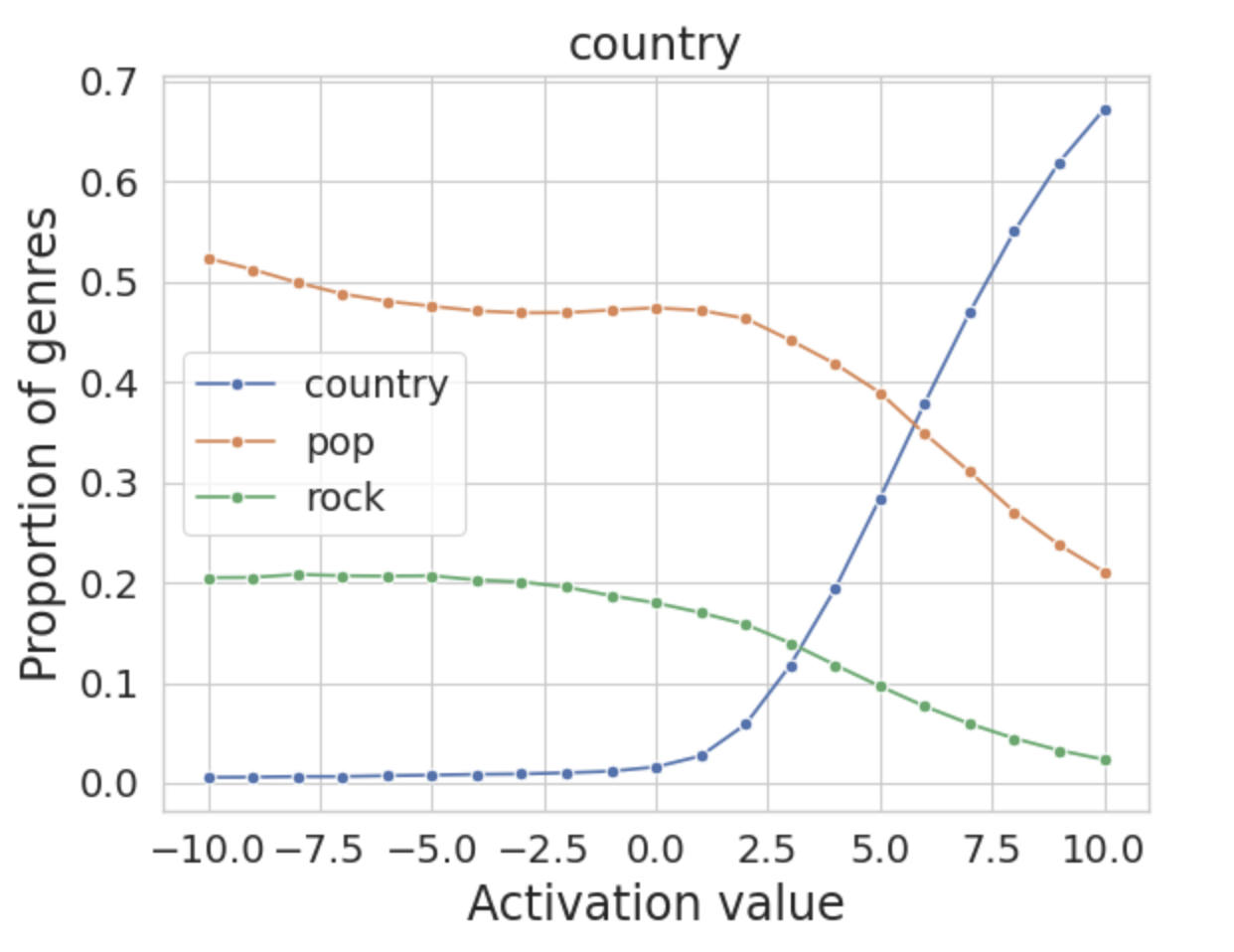}
 \end{subfigure}
 \begin{subfigure}{0.195\linewidth}
     \includegraphics[width=\textwidth]{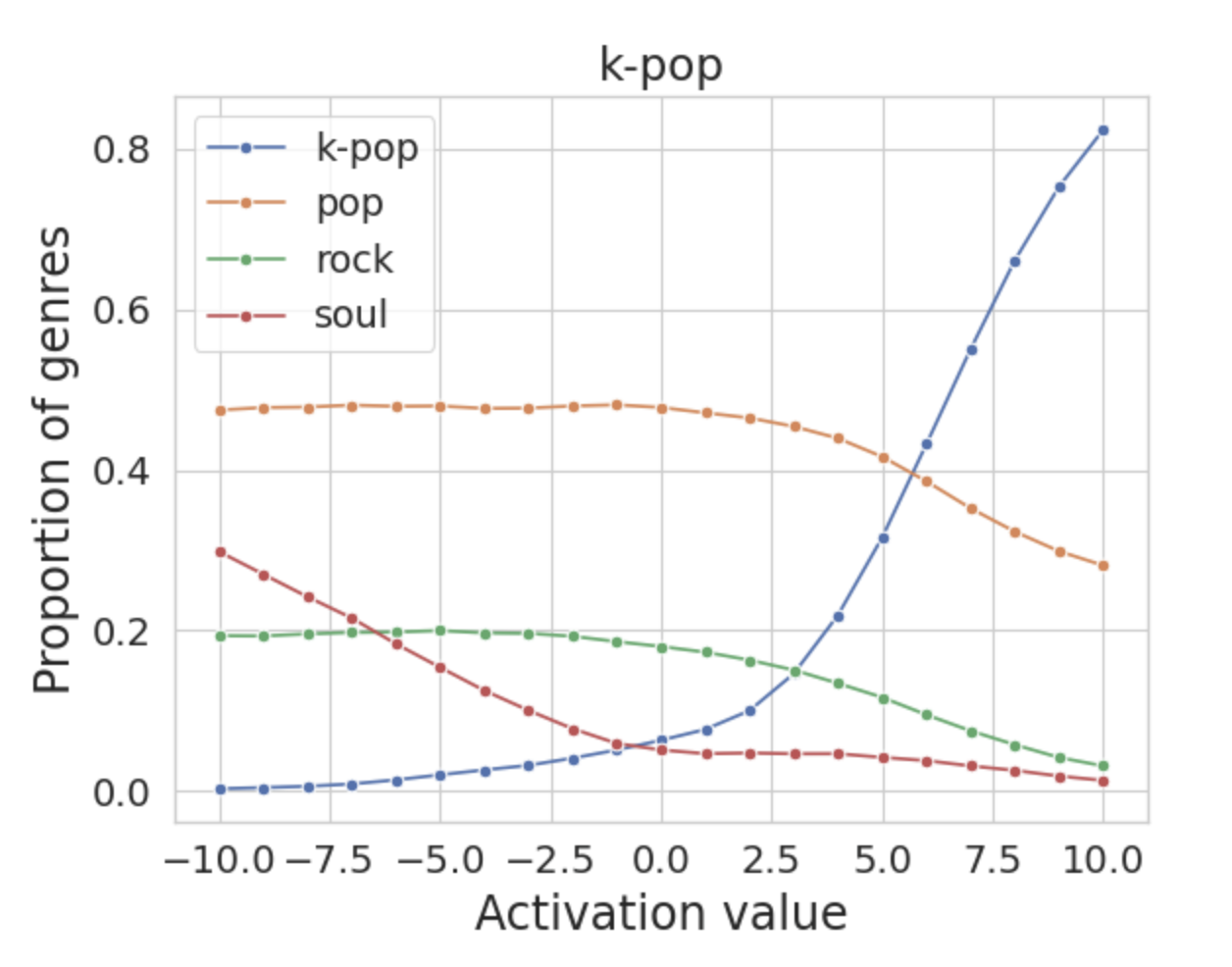}
 \end{subfigure}
 \begin{subfigure}{0.195\linewidth}
     \includegraphics[width=\textwidth]{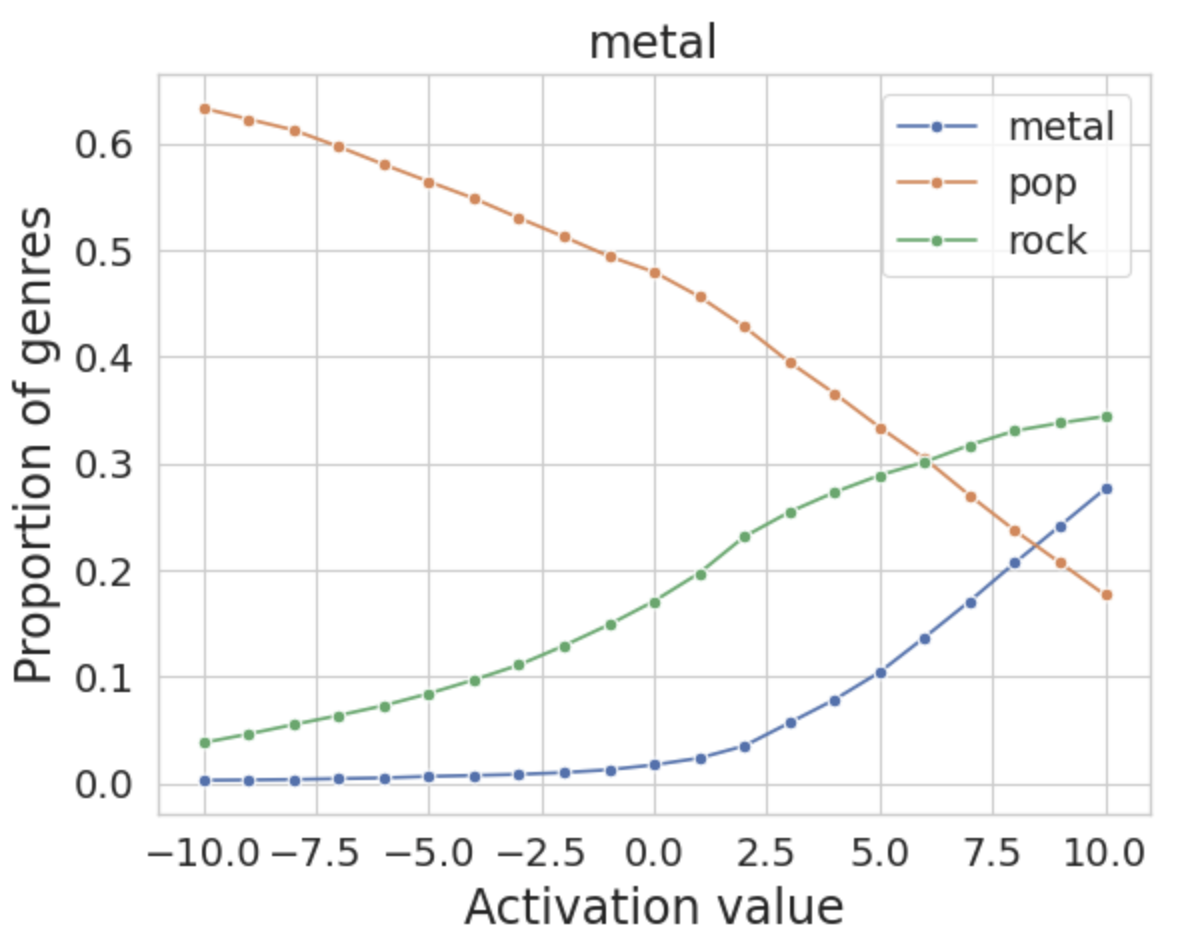}
 \end{subfigure}
 \begin{subfigure}{0.195\linewidth}
     \includegraphics[width=\textwidth]{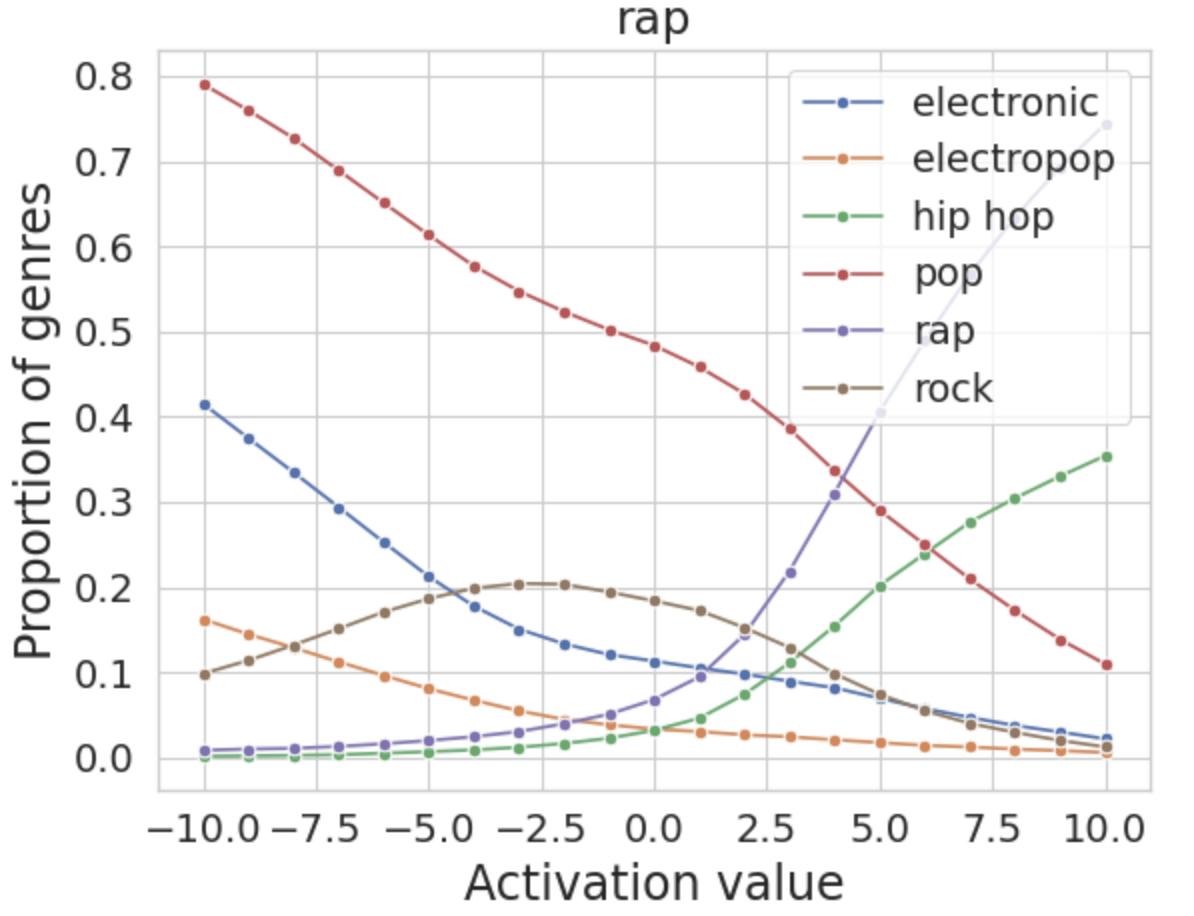}
 \end{subfigure}
 \begin{subfigure}{0.195\linewidth}
     \includegraphics[width=\textwidth]{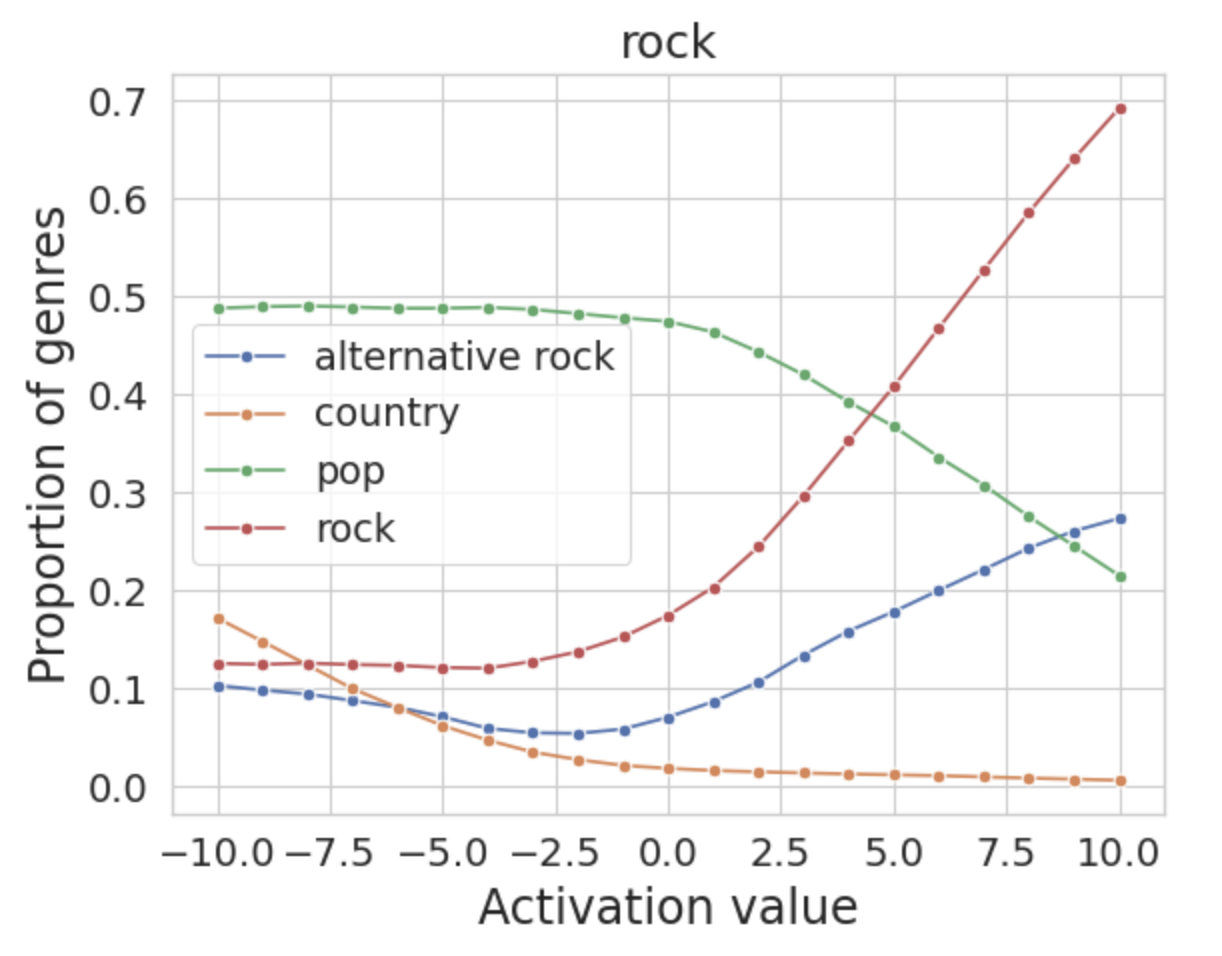}
 \end{subfigure}
 \caption{The proportion of genres in recommendations. Each plot corresponds to one single feature responsible for a given genre.}
 \end{subfigure}
 \begin{subfigure}{0.33\textwidth}
     \includegraphics[width=\textwidth]{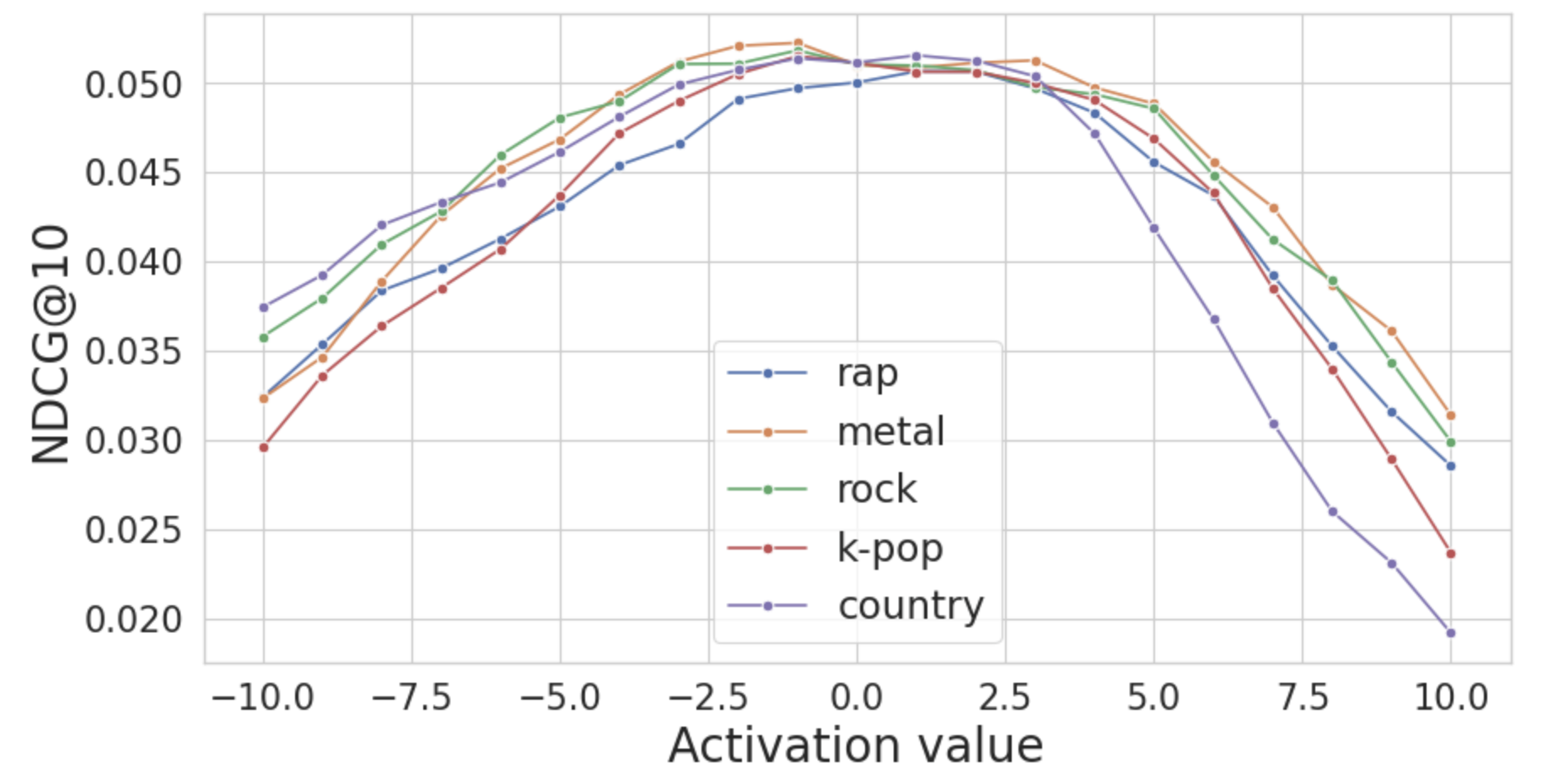}
     \caption{NDCG@10}
 \end{subfigure}
 \begin{subfigure}{0.33\textwidth}
     \includegraphics[width=\textwidth]{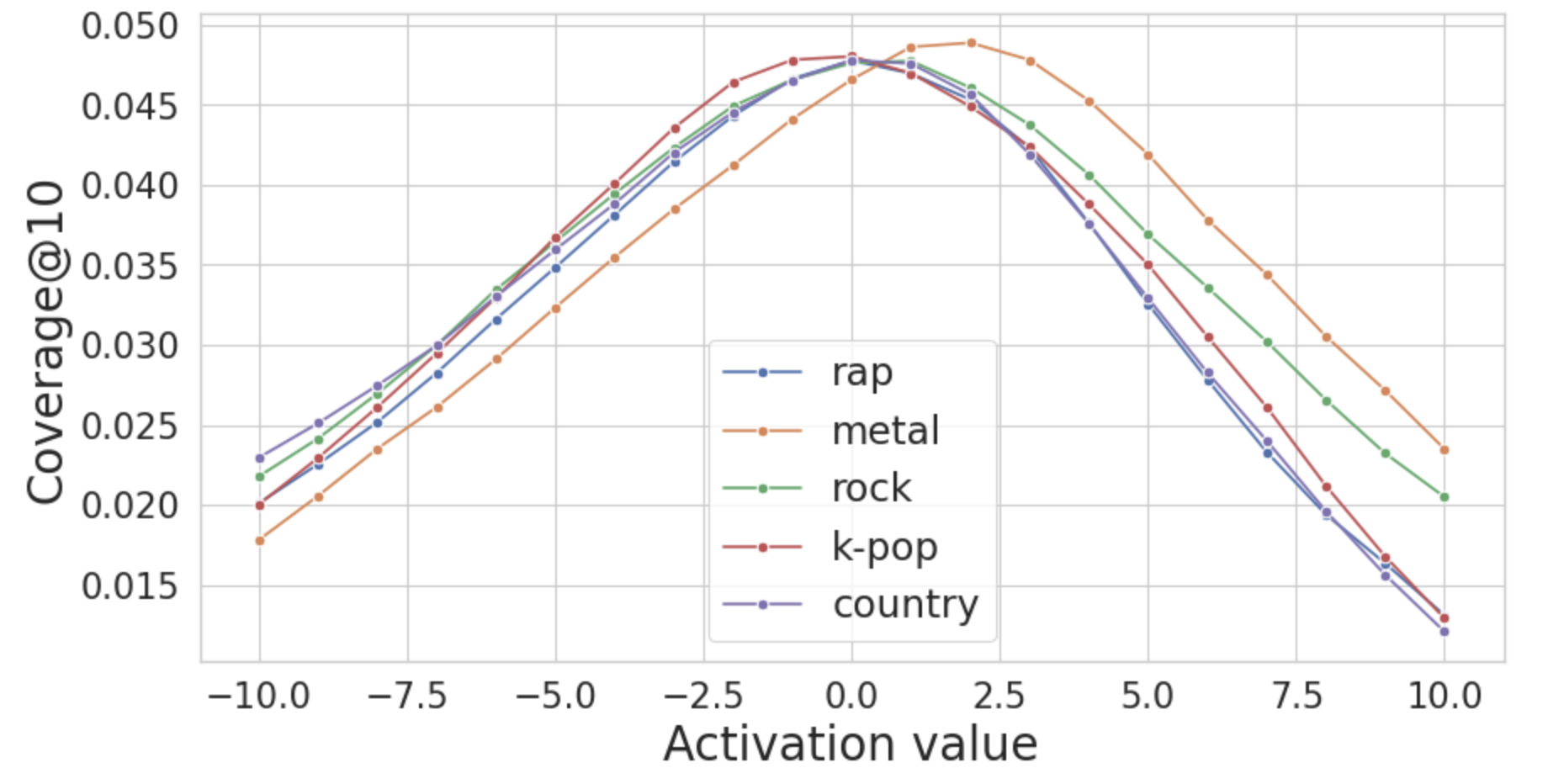}
     \caption{Coverage@10}
 \end{subfigure}
 \begin{subfigure}{0.33\textwidth}
     \includegraphics[width=\textwidth]{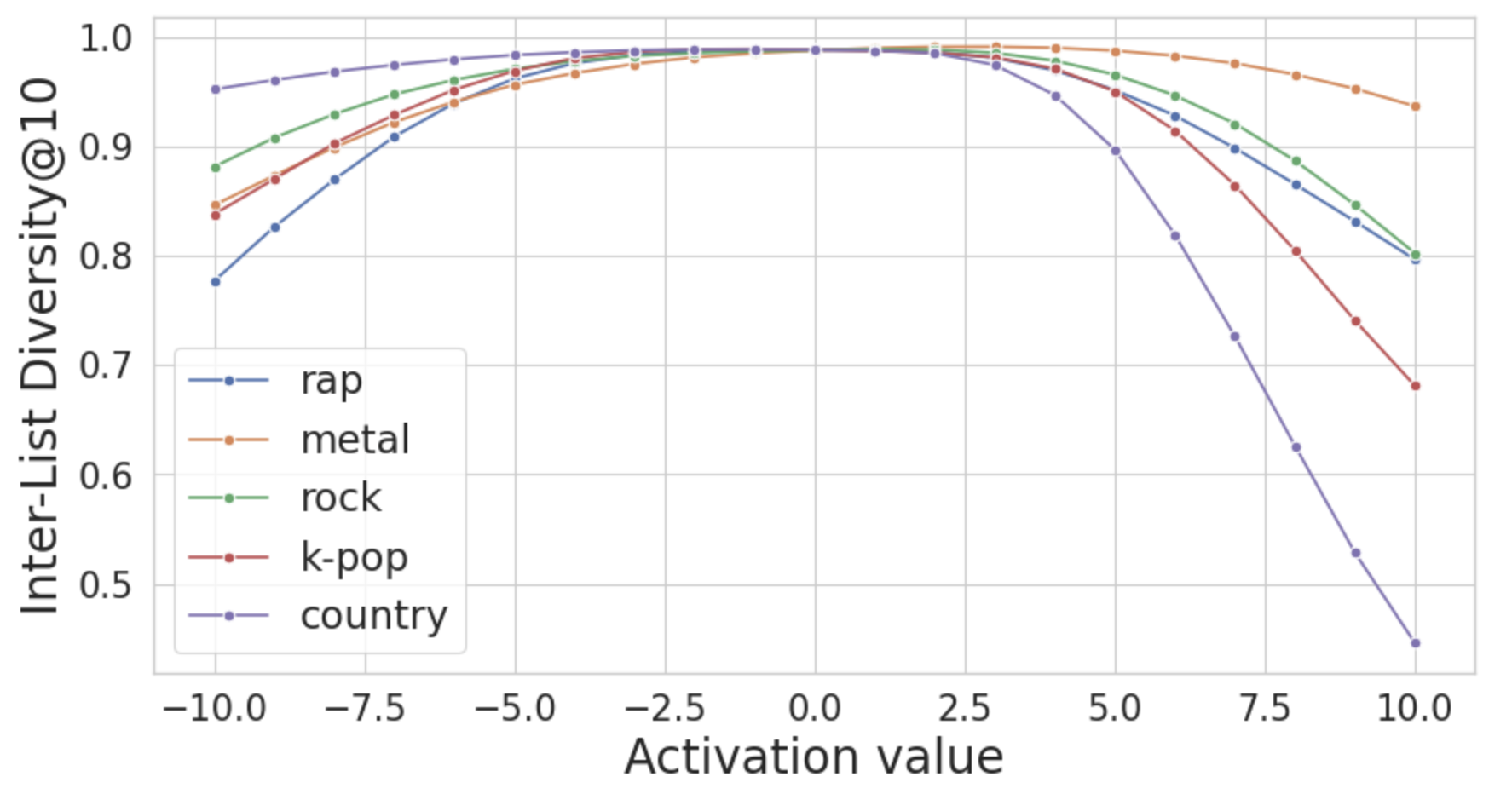}
     \caption{Inter-list Diversity@10}
     
 \end{subfigure}
\caption{Effect of changing a SAE feature on the proportion of genres in recommendations (a) and the recommendation metrics(b-d) for GPTRec on the Music4all dataset. For (a), each curve represents genre proportions for different activation values. Only genres with a large proportion of change are shown for clarity. For (b-d), each curve corresponds to one single feature that is responsible for a given genre.}
\label{fig:intervention_genres_music4all}

\end{figure*}

We also follow the procedure described in Section~\ref{sec:control_evaluation} and conduct extensive experiments on model control. Figure~\ref{fig:intervention_genres_music4all} illustrates how modifying a single SAE feature affects genre proportions and recommendation metrics. As with Movielens-20M, the proportion of the corresponding genre increases monotonically with activation value: it is near zero for large negative activations and reaches high levels for large positive activations. Additional patterns align with genre semantics and the correlations from Table 5. For example, increasing the rap feature also increases the proportion of hip-hop genre, increasing the metal feature increases the rock proportion, and increasing the rock feature boosts the proportion of alternative rock. In all these cases, the presence of pop genre tends to decrease.

The results regarding the effect on recommendation quality mirror previous findings: interventions with activation values below 2 lead to minimal changes in metrics, while stronger interventions can cause more significant degradation in performance.

\nopagebreak[4]
\begin{figure*}[!htb]
  \centering

 \begin{subfigure}{0.48\linewidth}
     \includegraphics[width=\textwidth]{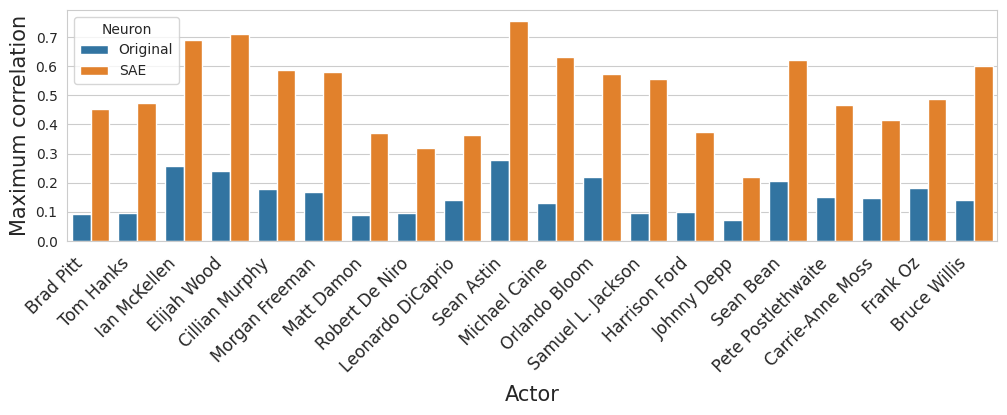}
     \caption{Actor}
     \label{fig:corr_original_vs_sae_actor}
 \end{subfigure}
 \begin{subfigure}{0.48\linewidth}
     \includegraphics[width=\textwidth]{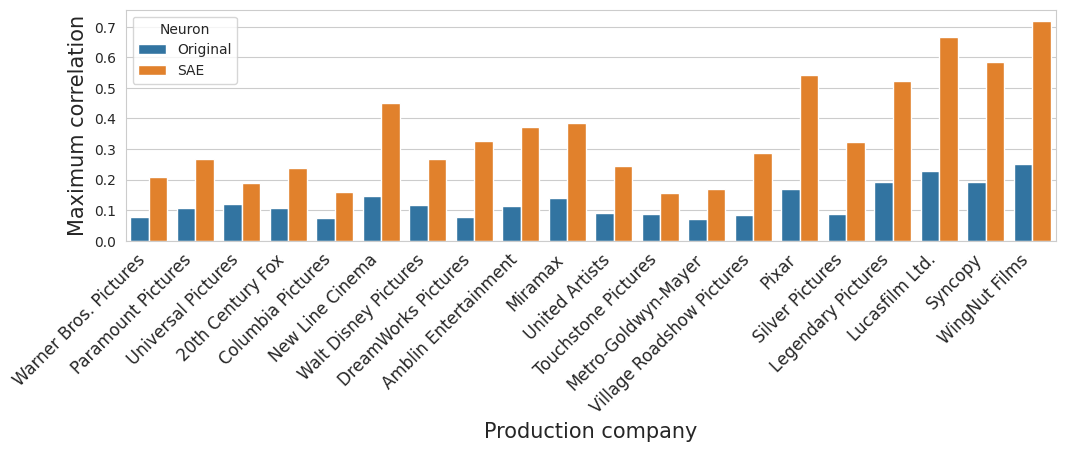}
     \caption{Company}
     \label{fig:corr_original_vs_sae_company}
 \end{subfigure}
  \begin{subfigure}{0.48\linewidth}
     \includegraphics[width=\textwidth]{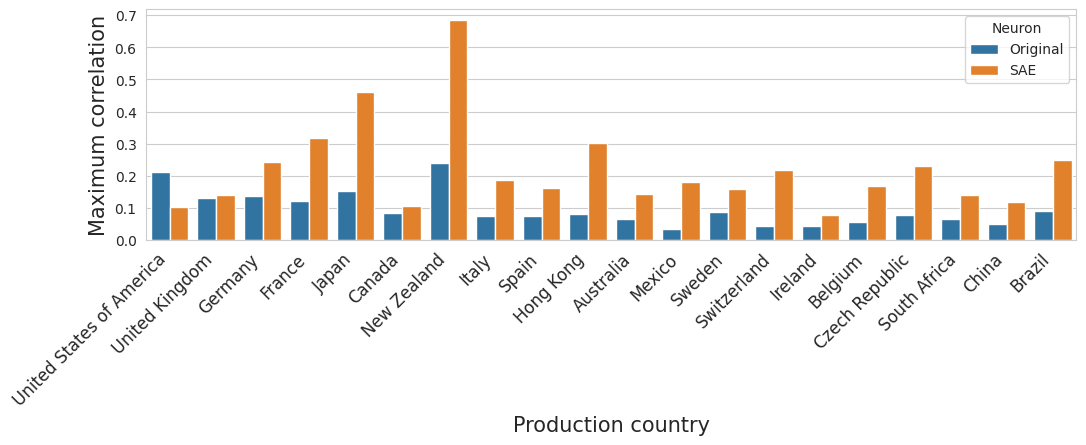}
     \caption{Country}
     \label{fig:corr_original_vs_sae_country}
 \end{subfigure}
  \begin{subfigure}{0.48\linewidth}
     \includegraphics[width=\textwidth]{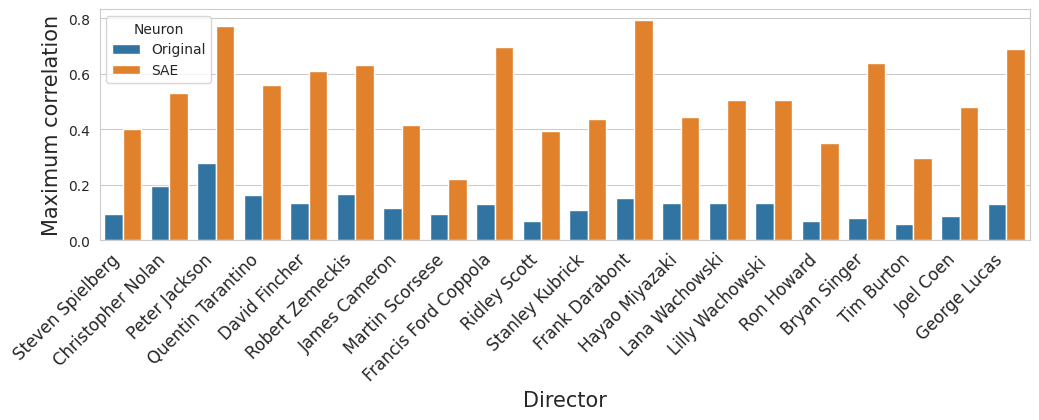}
     \caption{Director}
     \label{fig:corr_original_vs_sae_director}
 \end{subfigure}
 \caption{Comparison between interpretability of SAE features and neurons of the original transformer layer for different item attributes. The maximum correlation for each movie attribute is blue for the transformer layer and orange for SAE.}
 \label{fig:corr_original_vs_sae_attributes}
 
\end{figure*}

\nopagebreak[4]

\section{Analysis of Other Movie Attributes}
\label{sec:appendix_attributes}

To further explore the SAE feature space, we evaluate interpretability using additional item attributes beyond genres on the Movielens-20M dataset. Specifically, we consider the 20 most frequent actors, production companies, production countries, and directors. Following the procedure from Section~\ref{sec:original_vs_sae}, we compare the interpretability of SAE features with that of neurons from the original transformer layer.

Figure~\ref{fig:corr_original_vs_sae_attributes} shows the correlations between considered movie attributes and both the transformer neurons and SAE features. For all attributes, the association with SAE features is significantly stronger. Correlations are particularly high for actors and directors, moderate for production companies, and relatively low for most production countries.

Table~\ref{tab:sae_vs_transformer_other} summarizes the results, reporting the minimum, mean, and maximum correlation values for both transformer neurons and SAE features. These observations indicate that the SAE captures diverse and meaningful patterns beyond genre-level information.

\nopagebreak[4]

\begin{table}[t!]
\caption{Comparison of SAE features with original transformer layer neurons for different item attributes. Minimum, mean, and maximum correlation values for top features/neurons.}
\label{tab:sae_vs_transformer_other}
\centering
\small
\resizebox{0.8\columnwidth}{!}{
\begin{tabular}{@{}lrrrrrr@{}}
\toprule
\multirow{2}{*}{\textbf{Attribute}} & \multicolumn{3}{c}{\textbf{Transformer Layer}} & \multicolumn{3}{c}{\textbf{SAE Features}}   \\ 
\cmidrule(lr){2-4} \cmidrule(lr){5-7}
& \textbf{min}  & \textbf{mean}  & \textbf{max}  & \textbf{min} & \textbf{mean} & \textbf{max} \\
\midrule
Actor                             & 0.07          & 0.15           & 0.28          & 0.22         & 0.51          & 0.76         \\
Company                           & 0.07          & 0.13           & 0.25          & 0.15         & 0.35          & 0.72         \\
Country                           & 0.04          & 0.10           & 0.24          & 0.08         & 0.22          & 0.68         \\
Director                          & 0.06          & 0.13           & 0.28          & 0.22         & 0.52          & 0.79         \\ 
\bottomrule
\end{tabular}
}
\end{table}
\nopagebreak[4]

\end{document}